\documentclass[11pt,a4paper]{article}
\usepackage[latin1] {inputenc}
\usepackage{amsmath}
\usepackage{amsfonts}
\usepackage{amssymb}
\usepackage{enumerate}

\usepackage{tikz}
\usetikzlibrary{arrows,decorations.pathmorphing,backgrounds,fit}
\usepackage{pgfmath}
\usepackage{sidecap}

\newtheorem{definition}{Definition}
\newtheorem{prop}[definition]{Proposition}
\newtheorem{lemma}[definition]{Lemma}
\newtheorem{theorem}[definition]{Theorem}
\newtheorem{cor}[definition]{Corollary}
\newtheorem{obs}[definition]{Observation}

\newtheorem{claim}{Claim}

\newcommand{\Prob}[1]{\mathbf{Pr} \left( #1 \right)}
\newcommand\dw[2]{\draw[#1!#2,fill=#1!#2]}

\newcommand{\sC}{\mathcal{C}}
\newcommand{\sA}{\sc{A}}

\newcommand{\spat}{\mathfrak{s}}
\newcommand{\dest}{\mathfrak{d}}
\newcommand{\mmtm}{\mbox{\sc{Manhattan-mtm}}}

\newcommand{\sv}{\mbox{\sc{v}}}

\newcommand{\proof}{ \noindent \textit{Proof. }}
\newcommand{\qed}{\hspace{\stretch{1}$\square$}\bigskip}

\newcommand{\pset}{\mathcal{R}}

\newcommand{\tripset}{\mathcal{T}}

\newcommand{\tripout}[1]{\tripset^{\mathrm{out}}(#1)}
\newcommand{\tripin}[1]{\tripset^{\mathrm{in}}(#1)}

\newcommand{\trips}[2]{\tripset(#1, #2)}

\newcommand{\kernel}[1]{Ker(#1)}

\newcommand{\btrans}[1] {B_T^{#1}(i,j)}

\newcommand{\slow}[1] {\textsc{slow}_{#1}}
\newcommand{\slk} {\mathsf{slow}}
\newcommand{\crc} {\mathsf{cross}}

\newcommand{\wait} {\mathsf{pause}}

\newcommand{\bstop}[2] {B_{#1,k}^{#2}( i,j )}
\newcommand{\bcross}[2] {B_{C}^{#1,#2}( i,j )}

\newcommand{\sbg} { \sigma\mbox{-}B }
\newcommand{\sbtrans}[1] { \sigma\mbox{-}B_T^{#1} }
\newcommand{\sbstop}[2] {\sigma\mbox{-}B_{#1,k}^{#2}( i,j )}
\newcommand{\sbcross}[2] {\sigma\mbox{-}B_{C}^{#1,#2}}

\newcommand{\bpath}[1]{\mathsf{Bundle}(#1)}
\newcommand{\btripset}[1]{\tripset[#1]}
\newcommand{\btripout}[2]{\tripset[#1]^{\mathrm{out}}(#2)}
\newcommand{\btripin}[2]{\tripset[#1]^{\mathrm{in}}(#2)}

\title{\textbf{Modelling Mobility: A  \emph{Discrete} Revolution }
\\ {\small (Extended Abstract)} }

\author{Andrea   Clementi
\thanks{Contact Author: Dipartimento di Matematica, Universit\`a di Roma ``Tor Vergata''.
\texttt{E-Mail:~$\{$clementi$\}$@mat.uniroma2.it.}}
  \and Angelo Monti
\thanks{Dipartimento di Informatica, Universit{\`a}  di Roma
``La Sapienza". {\tt
E-Mail:~$\{$monti,silver$\}$@di.uniroma1.it.}}  \and Riccardo Silvestri$^\dagger$ }

\begin{document}

\maketitle

\begin{abstract}
We introduce a new approach    to model and analyze \emph{Mobility}. It  is fully based on discrete mathematics 
 and yields a
  class of mobility models, called  the \emph{Markov Trace} Model. This model 
   can be seen as the discrete version of the \emph{Random Trip} Model:
   including all variants of the
  \emph{Random Way-Point} Model \cite{L06}.

  We derive  fundamental properties and \emph{explicit} analytical    formulas for the     \emph{stationary distributions}   yielded by
   the Markov Trace Model.   Such results can be exploited to compute   formulas and properties for
   concrete  cases of the Markov Trace Model by just  applying     counting arguments.

     We    apply the above general results  to the discrete version
    of the \emph{Manhattan Random Way-Point} over a square   of bounded size.
    We get    formulas for the total stationary distribution and for two
    important \emph{conditional} ones: the agent spatial   and
      destination distributions.

     Our method makes the analysis of complex mobile systems  a feasible  task.
     As a further evidence of this important fact,  we first model  a complex  vehicular-mobile system over a 
     set of crossing streets. Several concrete issues  are   implemented such as parking zones, traffic lights, and
     variable vehicle  speeds.  By using a \emph{modular} version of the  Markov Trace Model, we get explicit formulas
     for the  stationary distributions yielded by this vehicular-mobile  model as well.
     
\end{abstract}

\bigskip
\noindent
\textbf{Keywords:} Models of Mobility,  Mobile Ad-Hoc Networks, Discrete Markov Chains.
 
 \bigskip
 

\thispagestyle{empty}

\newpage

\section{Introduction}
A crucial  issue in modeling mobility    is to find  a good balance between the goal of implementing  important
 features
of concrete  scenarios and the possibility to study the model    from an analytical point of view. Several
interesting approaches   have been introduced and studied over the last years
\cite{CBD02,DPSW08,LV05,L06}. Among them,  we focus on those models where   \emph{agents} move independently and
according to some random process, i.e., random mobility models. 
Nice examples of such random models are
the random-way point and the walker models \cite{CBD02,BRS03,DPSW08,L06,LV05} which are, in turn,
special cases of a  family of mobility models known as random trip model \cite{L06}. 

 Mobile networks are complex dynamical systems  whose    analysis   is far to be trivial. In particular, deriving explicit
formulas of the relative stationary probabilistic distributions, such  as the   agent spatial one,     
requires very complex integral calculus and/or sophisticated tools like the Palm Calculus
\cite{BB87,CNB04,L06,L07}.

 A first  goal of our  study is to make the analysis of such   dynamical  systems more accessible to 
 the (Theoretical) Computer Science   Community  by adopting  concepts and methods
  which are typical of this Community.   A possible  solution for this issue could be that of considering agents that
  walk over (random) paths of a graph. Nice results on random walks over graphs are   available \cite{AF99, DPSW08}, however,
  such models are not suitable to consider       agent speed variations, crossways, parking zones  and other concrete aspects
  of   mobile systems.

We   
 propose a new approach to model and analyse mobility. This approach is based on  a  simple observation
 over concrete network scenarios:

 \smallskip
 
\begin{quote}
 It is not so important to record  every position of
  the agents at every instant of time and it thus suffices to discretize the space
into a set of  cells and record  the current  agent cell at discrete  time steps.
\end{quote}
            
            \smallskip
           \noindent
            We exploit the above  observation    to get a class of fully-discrete mobility models based on
           \emph{ agent movement-traces} (in short,  traces).  A \emph{trace} is the representation of an  agent trajectory  by means of
           the  sequence of visited cells.  Similarly to the Random Trip Model, our mobile model is   defined by fixing 
           the set of feasible traces and the criterium the agent adopts to select
           the next    trace  after arriving at the end of the current trace.

           We   define  the   \emph{(Discrete) Markov Trace Model} (in short,  MTM) where, at every time step,
           an agent  either is  (deterministically) following the selected trace or  is  choosing at random (according to a given
           probability distribution) the next trace over a set of feasible traces, all starting from the final cell of the previous trace.
           
           \noindent It is important to observe that the same trajectory run at different speeds yields different traces (cell-sequences
           are in general ordered multi-sets): so, it is possible to  model   variable agent speeds that may depend on the specific 
           area traffic  or on other concrete issues.

A detailed description of  the  MTM is given in Section \ref{sec::MTM}.  We here discuss its  major features 
and  benefits.  

  Any MTM $\mathcal{D}$ determines a discrete-time  \emph{Markov chain}
$\mathcal{M}_{\mathcal{D}}$
  whose generic state   is   a   pair   $\langle T, i \rangle$: an agent has chosen trace $T$ and, at that time
  step, she is in  position $T(i)$.

  We first study the stationary distribution(s) of  the Markov chain $\mathcal{M}_{\mathcal{D}}$
     (in what follows we will   say shortly: ''stationary distribution of
  $\mathcal{D}$'').      
  We   show  evidence of the generality of our model,  derive an explicit  form of the stationary distribution(s), and 
 establish   \emph{existence} and \emph{uniqueness} conditions for the stationary distributions of an MTM.

           Our last result for general MTM is a necessary and sufficient condition for \emph{uniformness} of the
           stationary distribution of $\mathcal{D}$.

 \smallskip
 The above     results for the stationary phase can be applied to get explicit formulas for the stationary
 \emph{(agent) spatial distribution} and the stationary \emph{ (agent) destination} one.
 The former gives   the probability  that  an agent lies in a given cell, while the
 latter gives the probability that an agent, \emph{conditioned} to stay in a cell $v$, has \emph{destination} cell $w$,
  for any choice
 of $v$ and $w$.  
 
 The knowledge of such distributions is crucial  to  achieve   \emph{perfect simulation}, to derive  connectivity
 properties of  Mobile Ad-hoc  NETworkS (MANETS)  defined over  the mobility model, and for the 
 study of information spreading  over such MANETS  \cite{CMPS09,CPS09,CDMRV09}.

 We emphasize that all the obtained explicit formulas can be computed by   counting arguments  (it mainly concerns
 calculating the number of feasible traces passing over or starting from a cell).  If the    agent's behaviour
  can be described  by using a limited set of \emph{typical}  traces
  (this happens in most of MANETS applications), then such formulas can be computed by   a computer  in few minutes.

 Our  MTM model can thus  serve as a general framework that allows an analytical study of concrete mobility scenarios.
 We provide  two   examples  that show its   power and   applicability.

  In the first one,
 we consider the \emph{Manhattan Random Way-Point}  (MRWP) model \cite{CDMRV09,BRS03,L06}.
 This version of the   \emph{Random Way-Point} model    is motivated by  scenarios where 
 agents travel over an urban zone and try to
minimize the number of \emph{turns} while keeping the chosen route as short as possible.  
We then implement this model as a specific MTM and we     derive explicit formulas for  its    stationary distributions.
In particular, 
 we provide the spatial    and the destination
distributions for any choice of the cell resolution parameter $\epsilon>0$.
We observe that, by taking the limit for $\epsilon \rightarrow 0$, our explicit formula
of the  spatial distribution coincides to that computed by using   rather complex integral calculus
 in \cite{CDMRV09} for the continuous  space-time  MRWP model (in terms of \emph{probability density functions}).
 
 \noindent
 Finally,  we give, for the first
time, the destination distribution of the   continuous  space-time MRWP model as well.
 Both these  formulas  have been recently used to derive the first analytical  bounds on   flooding time for  the MRWP
model  \cite{CMS10b}.

Our approach can make  the analysis of complex scenarios much simpler:  it is just a matter of modelling
objects and events as   ingredients  of an MTM. After doing that, you do not  need to prove  new properties
or new formulas, you can just apply ours. 

\noindent
 As a second  concrete example of this fact,  we consider
a more complex vehicular-mobility  scenario:  The \emph{Down-Town} model 
where a set of horizontal and vertical
streets cross each other and they alternate with      building blocks (see Fig \ref{fig_downtown}).
  Agents (i.e.  vehicles) move over the streets 
according to \emph{Manhattan-like} paths and
park on the border of the streets  (a detailed description of the model is given in Section \ref{ssec::down}).
  Different agent speeds and
red and green  events of traffic lights    can be   implemented by considering  different traces over the same street path.

Thanks  to a  \emph{modular} 
version of our MTM model, we are also able to analyze this more
complex scenario. In fact, the main advantage of our approach is that a
given scenario can be analyzed by simply modelling objects and events
as  ''ingredient'' of an MTM, thus obtaining the stationary probability
distributions directly from our formulas.

\section{The Markov Trace Model} \label{sec::MTM}
The mobility model we are introducing is discrete with respect to time and space. The positions     an
agent can occupy during her movement belong to the set of \emph{points} $\pset$  and they are traced at discrete time steps. The set $\pset$ might be a subset of $\mathbb{R}^d$, for some $d =1,2,\ldots$, or it might be
some other set. It is only assumed that $\pset$ is a metric space.

\noindent A \emph{movement trace}, or simply a \emph{trace}, is any
(finite) sequence $T = (u_0,u_1,\ldots, u_k)$ of at least two points. When we mention points we tacitly assume that they
belong to $\pset$. The points of a trace are not necessarily distinct.
 The length of a trace $T$ (i.e., the number of points of $T$) is denoted by $|T|$ and,
for each $i = 0,1,\ldots |T|-1$, let $T(i)$ denote the $i$-th point of the trace $T$. 
A trace $T$ can be interpreted as the recording
of the movement of an agent starting from some initial time $t_0$: for every $i=0,1,\ldots,|T| - 1$,
$T(i)$ is the traced position of the agent at time $t_0 + i\cdot\tau$, where $\tau >0$ is the duration of a time step.

\noindent
In our model,
an agent can move along trajectories that are represented by traces.
For any trace $T$, let $T_{start}$ and $T_{end}$
denote, respectively, the starting point and the ending point of the trace. Let $\tripset$ be any set (possibly infinite) 
of traces.
We say that $\tripset$ is \emph{endless} if for every trace $T\in \tripset$, there is a trace $T'\in \tripset$ such
that $T'_{start} = T_{end}$.  

\noindent
For any
point $u\in\mathcal{R}$, $\tripout{u}$ denotes the subset of traces of $\tripset$ whose starting point is $u$.
Let \[ P(\tripset) = \{ u \;|\; \tripout{u} \neq\emptyset\} \ \mbox{ and } \
S(\tripset) = \{ \langle T, i \rangle \;|\; T\in \tripset \wedge 1 \leqslant i \leqslant |T| - 1\}   \]

\noindent
A \emph{Markov Trace Model} (\emph{MTM}) is a pair $\mathcal{D} = (\tripset, \Psi)$ such that:

 \noindent
 $i)$.
$\tripset$ is an endless trace set such that $|P(\tripset)| < \infty$;  

\noindent
$ii)$.
 $\Psi$ is a \emph{Trace Selecting Rule for} $\tripset$ (\emph{TSR}), that is,
$\Psi$ is a family of probability distributions $\{\psi_u\}_{u\in P(\tripset)}$ such that for
each point $u \in P(\tripset)$, $\psi_u$ is a probability distribution over $\tripout{u}$.

\noindent
Similarly to the random trip model \cite{L06}, 
any MTM $\mathcal{D}$ determines a Markov chain $\mathcal{M}_{\mathcal{D}} = (S(\tripset), P[\tripset, \Psi])$
whose state space is $S(\tripset)$ and the transition probabilities $P[\tripset, \Psi]$ are defined as follows:
 for every $T\in \tripset$, 
 
 \noindent
 - \emph{Deterministic-move Rule.} For every $i$ with $1 \leqslant i < |T| - 1$,
$ \Prob{\langle T, i \rangle \rightarrow \langle T, i + 1\rangle}  =  1$; 

\noindent
 - \emph{Next-trace Rule}. For every $T'$ in $\tripout{T_{end}}$,
$\Prob{\langle T, |T|-1\rangle \rightarrow \langle T', 1\rangle}  = \psi_{T_{end}}(T')$;  \\ 
all the other transition probabilities are 0.
It is immediate to verify that for any $s \in S(\tripset)$ it holds that
$\sum_{r \in S(\tripset)}  \Prob{s \rightarrow r}  =  1$.

\smallskip
\noindent
{\large \textbf{- Stationary Properties.}}
We first  introduce some notions that are useful in studying the stationary distributions of a MTM.
For any $u, v\in\pset$, $\trips{u}{v}$ denotes the subset of traces of $\tripset$ whose starting point is $u$ and whose
ending point is $v$.
A crucial Markov chain, determined by a MTM $\mathcal{D}$, is its
  \emph{Kernel}: its states are the   cells of $\mathcal{D}$, considered as \emph{turn} points
   (also known as \emph{way-points}),
  and the transition
  probability $\Prob{u \rightarrow v}$ equals the probability that an agent, lying on   $u$, chooses any
  trace ending in $v$.
Given a MTM $\mathcal{D} = (\tripset, \Psi)$, the \emph{Kernel of} $\mathcal{D}$ is the Markov chain
$\kernel{\mathcal{D}} = (P(\tripset), K[\tripset, \Psi])$ where the transition probabilities $K[\tripset, \Psi]$ are defined
as follows: for every $u, v\in P(\tripset)$,
{\small \[
\Prob{u \rightarrow v} \;=\; \left\{\begin{array}{ll}
\sum_{T\in \trips{u}{v}} \psi_u(T)      &    \mbox{if $\trips{u}{v} \neq \emptyset$}\\
0      &   \mbox{otherwise}
\end{array}\right.
 \]}
Observe that this definition is sound since, for every $u\in P(\tripset)$, it holds that
\[
\sum_{v\in P(\tripset)} \Prob{u \rightarrow v} \;=\;  \sum_{v\in P(\tripset)} \sum_{T\in \trips{u}{v}} \psi_u(T)
\;=\; \sum_{T \in \tripout{u}} \psi_u(T) \;=\; 1
\]
where the second equality derives from the fact that $\tripset$ is endless.
Along with a MTM $\mathcal{D} = (\tripset, \Psi)$ we will use the following notations. For every $u\in P(\tripset)$,
$
\Lambda_{\Psi}(u) =  \sum_{T\in \tripout{u}} (|T| - 1)\psi_u(T)$.
Observe that $\Lambda_{\Psi}(u)$ can be interpreted as the expected length of a trace starting from point $u$.
Define also 
$\tripin{u} = \{ T \in \tripset\;|\; T_{end} = u\}$.
The stationary distributions of $\mathcal{D}$ and that of $\kernel{\mathcal{D}}$ are strongly related as stated in the following

\begin{theorem}\label{kernel_stationary}
Let $\mathcal{D} = (\tripset, \Psi)$ be any MTM. The following properties hold.
$a)$
 A map $\pi : S(\tripset) \rightarrow \mathbb{R}$ is a stationary distribution of
$\mathcal{D}$ if and only if    
a stationary distribution $\sigma$ of $\kernel{\mathcal{D}}$ exists  such that
  \small \[
\forall \langle T, i\rangle \in S(\tripset) \qquad
\pi(\langle T, i\rangle) \;=\; \frac{1}{\sum_{u\in P(\tripset)} \sigma(u)\Lambda_{\Psi}(u)}\sigma(T_{start}) \psi_{T_{start}}(T)
\] \normalsize 
$b)$ A map $\sigma : P(\tripset) \rightarrow \mathbb{R}$ is a stationary distribution of
$\kernel{\mathcal{D}}$ if and only if
 a stationary distribution $\pi$ of $\mathcal{D}$ exists such that
\small \[
\forall u\in P(\tripset) \qquad
\sigma(u) \; =\; \frac{1}{\sum_{T\in \tripset} \pi(\langle T, 1\rangle)}\sum_{T\in \tripout{u}} \pi(\langle T, 1\rangle)
\] \normalsize
\end{theorem}



\noindent
\textbf{Stationary Distributions: Existence and Uniqueness.}

\begin{cor}\label{kernel_stationary_cor}
Let $\mathcal{D} = (\tripset, \Psi)$ be any MTM. The following properties hold.
$a)$
$\mathcal{D}$ has always a stationary distribution.
$b)$
$\mathcal{D}$ has a unique stationary distribution if and only if
 $\kernel{\mathcal{D}}$ has a unique stationary distribution.
\end{cor}

\noindent
Next proposition shows that the Kernel of an MTM can be any finite Markov chain.

\begin{prop}\label{kernel_general}
Given any  Markov chain $\mathcal{M} = (S, P)$ with $S \subseteq \pset$, there exists a MTM $\mathcal{D}$
such that $\kernel{\mathcal{D}} = \mathcal{M}$.
\end{prop}

\noindent Let $\mathcal{D} = (\tripset, \Psi)$ be a MTM. For any two distinct points $u, v\in P(\tripset)$, we say that
$u$ is \emph{connected to} $v$ in $\mathcal{D}$ if there exists a sequence of points of $P(\tripset)$
$(z_0, z_1,\ldots,z_k)$ such that $z_0 = u$, $z_k = v$, and, for every $i = 0,1,\ldots, k-1$,
$
\sum_{T\in \trips{z_i}{z_i+1}} \psi_{z_i}(T) > 0
$
Informally, this can be interpreted as saying that if an  agent is
in $u$ then, with positive probability, she will reach $v$. We say that $\mathcal{D}$ is \emph{strongly
connected} if, for every $u, v\in P(\tripset)$, $u$ is connected to $v$. Observe that if $\mathcal{D}$ is not strongly
connected then     at least a pair of points $u,v\in P(\tripset)$ exists such that $u$ is not connected to $v$.

\begin{theorem}\label{uniqueness}
If $\mathcal{D}$ is a strongly connected MTM then $\mathcal{D}$ has a unique stationary distribution.
\end{theorem}

\noindent
{\bf Stationary Distributions: Uniformity.}
Let $\mathcal{D} = (\tripset, \Psi)$ be any MTM. We say that $\mathcal{D}$ is \emph{uniformly selective} if
$\forall u\in P(\tripset)$, $\psi_u$ is a uniform distribution.
We say that $\mathcal{D}$ is \emph{balanced} if
$\forall u\in P(\tripset)$,  $|\tripin{u}| \;=\; |\tripout{u}|$
Observe that if a MTM $\mathcal{D} = (\tripset, \Psi)$ has a uniform stationary distribution then it must be the case that
$|S(\tripset)| < \infty$ or, equivalently, $|\tripset| < \infty$.

\begin{theorem}\label{uniformity}
A MTM $\mathcal{D} = (\tripset, \Psi)$ has a uniform stationary distribution if and only if it is both
uniformly selective and balanced.
\end{theorem}

  \noindent
  {\bf Stationary Spatial and Destination Distributions.} 
We use the following notations.  For any trace $T \in  \tripset$ and for any $u \in \pset$, define 

\small \[  \#_{T,u}  =   | \{  i \in \mathbb{N} \  | \ 1 \leqslant  i  <  |T|-1  \ \wedge \ T(i) = u \}|    
   \  \mbox{ and } \ \tripset_u = \{ T \in \tripset \ | \ \#_{T,u} \geqslant 1 \} \] \normalsize

\noindent
- We now     derive the function $\spat(v)$   representing the probability that an agent lies  in point 
 $v\in \pset$
w.r.t. the   stationary distribution $\pi$.  This is called  \emph{Stationary  (Agent) Spatial  Distribution}.
 By definition,  for any point $u \in\pset $,  it holds that

\small \[
  \spat(u)  =   \sum_{\langle T,i \rangle \in  S(\tripset)  \wedge T(i) = v} \pi(\langle T,i\rangle) 
    =  \sum_{T \in \tripset_u}  \#_{T,u} \cdot \pi(\langle T,1\rangle)  \label{pos1}
\] \normalsize

\noindent If  the   stationary distribution $\pi$  is uniform, then     
$
\spat(u) \ =  \ (1 / | S(\tripset) | )  \cdot  \sum_{T \in \tripset_u}  \#_{T,u}$. \\
We   say that an MTM is \emph{simple} if,  for  any trace   $T \in \tripset$ and $u \in \pset$,  
$\#_{T,u} \leqslant 1$.   Then, if   the MTM is simple \emph{and}   $\pi$   is uniform, then it holds 

\begin{equation}\label{pos}
 \spat(u) \ =  \ \frac {|\tripset_u|} {| S(\tripset) |}  
\end{equation}

\noindent
- Another important   distribution is given by    function $\dest_u(v)$ representing
 the   probability 
 that an agent has destination $v$ under the condition she  is in position $u$.
   This function will be called \emph{Stationary  (Agent) Destination Distribution.}  By definition, it holds that
{\small \[
  \dest_u(v)   =   
   \frac{\sum_{\langle T,i \rangle \in  S(\tripset)  \wedge T(i) = u \wedge T_{end} = v} \pi(\langle T,i\rangle) }{ \spat(u)  }   
    =  \frac{\sum_{   T \in \tripset_u \wedge T_{end} = v} \#_{T,u} \cdot \pi(\langle T,1\rangle) }{ \spat(u)  } 
       \]}   
       
       \noindent
     We define  $ \Gamma_u(v) = | \tripset_u \cap \tripin{v}|$
          and $\Gamma_u = | \tripset_u |$ and 
 observe that, if   the MTM is simple \emph{and}   $\pi$   is uniform, then  
   {\small  \begin{equation}\label{eq::destsimple}  
   \dest_u(v) =   \frac{ \Gamma_u(v)  }{ \Gamma_u}    \end{equation}
}

\section{The Manhattan Random-Way Point} \label{sec::Manhattan}

In this section, we study  a mobility model, called \emph{Manhattan Random-Way Point}, an
interesting variant of the Random-Way Point that has been recently studied  in \cite{CDMRV09}. \\
Consider a finite 2-dimensional square of edge length $L>0$. A set $\sA$ of $n$ independent \emph{Agents} move
over this square according to the following random rule. Starting from an initial   position $(x_0,y_0)$,  every
agent  selects a   \emph{destination}
 $(x,y)$ uniformly at random
in  the square (i.e. every point of the square has the same probability to be chosen). Then, the agent chooses
(again \emph{uniformly at random})  between  the two    feasible \emph{Manhattan (shortest) paths}. 
  Once the   destination and the feasible path are randomly selected,   the agents start  following  the chosen route
 with  \emph{speed} determined by the parameter $\sv$.
  In the sequel, we  assume that all agents have the same speed $\sv$ that represents
     the travelled distance by an agent in the time unit. However,  a variable agent speed can be easily
     modelled and analyzed  by considering  more traces for  any\emph{ source-destination} pair.
     Once arrived at the selected destination, every agent  independently re-applies the
      process described above again and again. This  
     infinite process yields the \emph{Manhattan Random Way-Point}.

     \noindent
     We  here    consider a discrete version of  the Manhattan Random-Way Point which is  in fact a   \emph{Markov Trace Model}.
     Agents will act over a \emph{square
     cell grid } of \emph{arbitrary-high resolution} according to a global discrete clock.
     Every agent, within the next time step,
      can reach any grid point (that can be also considered as  a square cell)  which is adjacent to its current position.   
       We emphasize that, as the grid
      resolution increases, the time unit    decreases.
      This scalability allows to observe the process at arbitrary-small time unit and space resolution
      while preserving the ability to choice any possible agent speed.

      \noindent
      In order to formally define the   MTM, we introduce the following \emph{support  
    graph}  $G_{\epsilon}(V_{\epsilon},E_{\epsilon})$  where 
    $
V_{\epsilon} = \{ (i\epsilon, j\epsilon) \,  : \, i,j\in \{0,1,\ldots, N-1\}\}$   and
 $
E_{\epsilon} = \{ (u,v) \, : \, u, v \in V_{\epsilon}\;\; \wedge \ d(u,v) = \epsilon \}$
where, here and in the sequel,   $N = \lceil L / \epsilon \rceil$ and  $d(\cdot, \cdot)$ is the Euclidean
distance.


\noindent  Now,  given any point  $v
\in V_{\epsilon}$,  we define a set $\sC (v)$ of feasible paths from $v$ as follows. For any   point $u$,  
  $\sC(v)$ includes  the (at most) two Manhattan paths having exactly one corner point. More precisely,  let $v =
(x,y)$ and $u =( x', y')$ we consider the path having first the horizontal segment from $(x,y)$ to $(x',y)$ and
then the vertical segment to $(x',y')$. The second path is symmetrically formed by the vertical segment from
$(x,y)$ to $(x,y')$ and the horizontal segment to $(x',y')$. Observe that if $x = x'$ or $y = y'$, then the two
paths coincides.  We are now able to define the \emph{Manhattan Markov Trace Model} (in short $\mmtm$)
$(\tripset_{\epsilon},\Psi_{\epsilon})$, where \\
$ \tripset_{\epsilon} = \{ T \  |   \  \text{ $T$ is the point sequence of a path in $\sC(v)$ for some $v \in V_{\epsilon}$ }  \}$, \\
and $\Psi_\epsilon$ is the \emph{uniform} TSR for $\tripset_{\epsilon}$.  It is easy to verify   the $\mmtm$ enjoys the following properties.
 
\begin{obs} \label{obs::revers}
 The $\mmtm$ is balanced, uniformly-selective and strongly-connected. So, from Theorems \ref{uniformity}
 and \ref{uniqueness}, the $\mmtm$ has a unique stationary distribution and it is the uniform one.
 Moreover, since the $\mmtm$ is  simple, the stationary spatial and the destination distributions are given by
 Eq.s  \ref{pos} and \ref{eq::destsimple}, respectively.
 \end{obs}

\noindent
So, we just have to count the size of some  subsets of  traces (i.e paths in $G_{\epsilon}(V_{\epsilon},E_{\epsilon})$).
     Due to lack of space,  all calculations are given in App. \ref{app::manhatdest}.   
    The point $(i\epsilon,j\epsilon)$ will be denoted by its grid
 coordinates $(i,j)$.

\noindent  The stationary spatial distribution for  $(\tripset_{\epsilon},\Psi_{\epsilon})$ is 
{\small  \begin{equation}\label{posm}
\spat_\epsilon (i,j)=\frac{3( (4N^2-6N+2)(i+j)-(4N-2)(i^2+j^2)+6N^2-8N+3)}{(N^4-N^2)(4N-2)} 
\end{equation} }

\begin{SCfigure}
\begin{tikzpicture}
[scale=0.6]
\def\LL{10}
\pgfmathsetmacro{\LLQ}{200/(\LL*\LL)}
\def\list{0,.2,...,\LL}
\foreach \x in \list {
    \foreach \y in \list {
        \pgfmathsetmacro{\d}{(\LLQ*(\LL*(\x + \y) - \x*\x - \y*\y))}
        \dw{black}{\d} (\x, \y) +(-.1, -.1) rectangle ++(.1, .1);
    }
}
\def\dest{blue}
\pgfmathsetmacro{\NN}{100/\LL}
\pgfmathsetmacro{\XX}{floor((\LL/0.2)/3)*0.2}
\pgfmathsetmacro{\YY}{floor((\LL/0.2)/4)*0.2}
\pgfmathsetmacro{\d}{\NN*\YY}
\pgfmathsetmacro{\LY}{\LL + .1 - \YY}
\dw{\dest}{\d} (\XX, \YY) +(-.1, -.1) rectangle ++(.1, \LY); 
\pgfmathsetmacro{\d}{\NN*(\LL - \YY)}
\pgfmathsetmacro{\YYZ}{\YY + .1}
\dw{\dest}{\d} (\XX, \YY) +(-.1, -\YYZ) rectangle ++(.1, .1); 
\pgfmathsetmacro{\d}{\NN*(\LL - \XX)}
\pgfmathsetmacro{\XXZ}{\XX + .1}
\dw{\dest}{\d} (\XX, \YY) +(-\XXZ, -.1) rectangle ++(.1, .1); 
\pgfmathsetmacro{\d}{\NN*\XX}
\pgfmathsetmacro{\LX}{\LL + .1 - \XX}
\dw{\dest}{\d} (\XX, \YY) +(-.1, -.1) rectangle ++(\LX, .1); 
\dw{\dest}{50} (\XX, \YY) +(-.1, -.1) rectangle ++(.1, .1); 
\end{tikzpicture}
\caption{The spatial  density function   is shown by a gradation 
of gray (black corresponds to the maximum density and white corresponds to the 
minimum density).  The destination probability over the cross 
of   agent   position $(L/3, L/4)$ is shown in gradation of blue.}   \label{fig_manhattan}
\end{SCfigure}
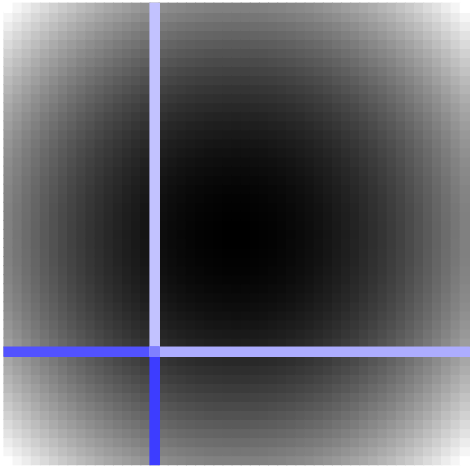

\noindent
 We now study the  Manhattan Random-Way Point  over grids of arbitrarily high
\emph{resolution}, i.e. for $\epsilon \rightarrow 0$ in order to     derive the \emph{probability densitiy}
functions of the   stationary distributions.  We first compute the  probability that an agent lies into a square of
center $(x,y)$ (where $x$ and $y$ are the Euclidean coordinates of a point in $V_{\epsilon}$) and side length $2
\delta$ w.r.t.  the   spatial  distribution; then, we take the limits as $\delta,\epsilon \rightarrow 0$.  We thus get 
 the     \emph{probability
density function} of the spatial distribution (see Fig. \ref{fig_manhattan})   
{\small    \begin{equation}\label{rpd}
s(x,y) = \frac{3}{L^3}(x+y)-\frac{3}{L^4}({x}^2+{y}^2)
\end{equation} }


\noindent

\noindent
This is also the formula obtained in \cite{CDMRV09}  for the classic  
 (real time-space) MRWP model.

\noindent  
 \normalsize 
The   stationary destination density function can be computed very similarly  by applying Eq. \ref{eq::destsimple}.
This is described in Appendix \ref{app::manhatdest}. 
The probability $f_{(x_0,y_0)}(x,y)$
that an agent, conditioned to stay in position $(x_0,y_0)$,
has destination $(x,y)$ is 
{\footnotesize 
\begin{equation} \label{eq::DENSITY}
 f_{(x_0,y_0)}(x,y) =\left\{
\begin{array}{ll}
  \frac {2L - x_0 - y_0}{4L(L(x_0+y_0) -(x_0^2 +y_0^2 ))} & \mbox{if $x<x_0$ and $y<y_0$}\\
\frac { x_0+y_0} {4L(L(x_0+y_0) -(x_0^2 +y_0^2 ))}     & \mbox{if $x>x_0$ and $y>y_0$}\\
     \frac { L-x_0 + y_0 }{4L(L(x_0+y_0) -(x_0^2 +y_0^2 ))} & \mbox{if $x<x_0$ and $y>y_0$}\\
  \frac { L + x_0 -y_0}{4L(L(x_0+y_0) -(x_0^2 +y_0^2 ))} & \mbox{if $x>x_0$ and $y<y_0$}\\
 + \infty  & \mbox{if $x=x_0$ and $y=y_0$} \\
  + \infty& \mbox{if $x=x_0$ and $y<y_0$  (South Case)}\\
+ \infty  & \mbox{if $x<x_0$ and $y=y_0$  (West  Case}\\
  + \infty & \mbox{if $x=x_0$ and $y>y_0$  (North Case)}\\
 + \infty & \mbox{if $x>x_0$ and $y=y_0$  (East Case)}\\
\end{array}
\right.
\end{equation}
} 
It is also possible to derive the probability  that an agent, visiting point $(x_0,y_0)$,
has destination in one of the last four cases (south, west, north, and east) (see Fig. \ref{fig_manhattan})
{\footnotesize \begin{eqnarray*} 
 \phi^{\mbox{south}}_{(x_0,y_0)}   =  \phi^{\mbox{north}}_{(x_0,y_0)} =
\frac{y_0(L-y_0)}{4L(x_0+y_0) - 4(x_0^2 + y_0^2)} \ , & & 
 \phi^{\mbox{west}}_{(x_0,y_0)}  = \phi^{\mbox{east}}_{(x_0,y_0)}   =
\frac{x_0(L-x_0)}{4L(x_0+y_0) - 4(x_0^2 + y_0^2)} \   
 \end{eqnarray*}
}
\noindent
We observe  that    the resulting   \emph{cross} probability, (i.e.  the probability   
  an agent has destination over the \emph{cross} centered on its current position),  is equal to $1/2$ despite the
 fact that this region (i.e. the cross) has area 0. This is  crucial for getting an upper bound on flooding time
 \cite{CMS10b}.

\section{Modular Trace Models} \label{sec::modular}

Defining a MTM whose aim is the approximate representation of a concrete  mobility scenario might be
a very demanding task. We thus  introduce a technique that makes 
the definition of MTMs easier  when the mobility scenario is \emph{modular}. For example, consider vehicular mobility 
in a city. Any mobility trace can be viewed as formed by the concatenation of \emph{trace segments} each of which is the 
segment of the trace that lies on a suitable segment of a street (e.g., the segment of a street between two 
crossings). Moreover, given a street segment we can consider all the trace segments that lies on it. Then, it is reasonable
to think that two alike street segments (e.g., two rectilinear segments approximately of the same length),
have similar collections of trace segments. This leads us to the insight that all the traces can be defined by
suitably combining the collection of trace segments relative to street segments. It  works just like   combining  
\emph{Lego} blocks.

\noindent
In the sequel, we use the term \emph{trace segment} to mean a trace that is a part of longer traces.
Given a trace (or a trace segment) $T$, the \emph{shadow of} $T$, denoted by $S_T$, is the sequence of points
obtained from $T$ by replacing each maximal run of repetitions of a point $u$ by a 
single occurrence of $u$. For example, the shadow of $(u, u, v, w, w, w, u)$ is $(u, v, w, u)$ (where $u$, $v$, $w$ are
distinct points). 
Given any two sequences of points $T$ and $T'$ (be them traces, trace segments, or shadows), 
the \emph{combination} of $T$ and $T'$, in symbols $T\cdot T'$, is the concatenation 
of the two sequences of points. Moreover, we say that $T$ and $T'$ are \emph{disjoint} if no point occurs
in both $T$ and $T'$. 
Given any multiset $X$ we denote  the cardinality of $X$ 
by $|X|$, including repeated memberships.

\noindent
A \emph{bundle} $B$ is any non-empty finite multiset of trace segments such that, for every $T, T'\in B$,
$S_T = S_{T'}$. The common shadow of all the trace segments in $B$ is called the \emph{shadow of}
 $B$ and it is denoted by $S_B$. 
\\  Two bundles $B$ and $B'$ are \emph{non-overlapping} if $S_B$ and
 $S_{B'}$ are disjoint. Given two non-overlapping bundles $B$ and $B'$ the \emph{combination} of $B$
 and $B'$, in symbols $B\cdot B'$, is the bundle consisting of all the trace segments $T\cdot T'$ for all the 
 $T, T'$ with $T\in B$ and $T'\in B'$. Notice that, since $B$ and $B'$ are non-overlapping, it holds that
 $S_{B\cdot B'} = S_B\cdot S_{B'}$. Moreover, it holds that $|B\cdot B'| = |B|\cdot |B'|$.

\noindent
 A \emph{bundle-path} is a sequence of bundles $P = (B_1,B_2,\ldots,B_k)$ such that any two consecutive 
 bundles of $P$ are non-overlapping. A bundle-path $P$ determines a bundle $\bpath{P} = B_1\cdot B_2\cdots B_k$.
 Observe that  
 $ |\bpath{P}| \;=\; \prod_{i = 1}^k |B_i|$.
Given a bundle-path $P$, let $P_{start}$ and $P_{end}$ denote, respectively, the starting point and the ending 
point of the traces belonging to $\bpath{P}$.

\noindent
A {\em route} $R$ is a multiset of bundle-paths all having the same starting point $R_{start}$ and 
the same ending point $R_{end}$ (i.e. there exist points $R_{start}$ and $R_{end}$ such that for every bundle-path $P$ in $R$ it holds 
$P_{start}=R_{start}$ and $P_{end}=R_{end}$).

 \noindent
  Informally speaking,  a bundle-path is formed by traces having the same shadow; introducing such  different  traces allows to  model agents travelling on the same
path at different speeds (in this way, it is also possible to change speed around cross-ways and modeling other concrete events).
Moreover, routes are introduced to allow different bundle-paths connecting two points. By introducing more copies of the same bundle-path into a route, it is possible
to determine   paths having more agent traffic.  As described below, all  such issues can be implemented  without making the system analysis much harder: it  
still mainly concerns counting traces   visiting  a given bundle.

\noindent
A \emph{Route System} is a pair $\mathfrak{R} = (\mathcal{B}, \mathcal{R})$ where: 
$(i)$
$\mathcal{B}$ is a set of bundles, and
$(ii)$
$\mathcal{R}$ is a multiset of routes over the bundles of $\mathcal{B}$ such that, 
for every $R\in \mathcal{R}$, there exists $R'\in \mathcal{R}$ with $R'_{start} = R_{end}$.

\noindent
We need some further notations. 
Let $\mathcal{R}_u$ be the multiset $\left\{ R\in \mathcal{R} |R_{start}=u  \right\}$.
  $\#_{P,T}$ is the multiplicity of trace $T$ in $\bpath{P}$.
 $\#_{P,B}$ is the number of occurrences of bundle $B$ in the bundle-path $P$.
Moreover $\#_{R,B}=\sum_{p\in R}\#_{P,B}$ and $\#_{B,u}=\sum_{T\in B}\#_{T,u}$, 
where the sums vary over all the elements, including repeated memberships.
 Let $\#B$ denote the total number of occurrences of points in all the trace segments of $B$, including repeated
 memberships, that is,
 {\small \[
\#B \;=\;  \sum_{\text{$u$   in $S_B$}}\#_{B, u}
 \]}
A Route System $\mathfrak{R} = (\mathcal{B}, \mathcal{R})$ defines a MTM $\mathcal{D}[\mathfrak{R}] =
(\btripset{\mathfrak{R}}, \Psi[\mathfrak{R}])$ where \\
$(i)$
$\btripset{\mathfrak{R}} \;=\; \{ T \;|\; \exists R\in \mathcal{R} \,\,\exists P\in R :\; T \in \bpath{P} \}
$
Notice that $\btripset{\mathfrak{R}} $ is a set not a multiset.
$(ii)$
for every $u\in P(\btripset{\mathfrak{R}})$ and for every $T \in \btripout{\mathfrak{R}}{u}$, 
{\small \[
\psi[\mathfrak{R}]_u(T) \;=\; \frac{1}{\left|\mathcal{R}_u\right|}\sum_{R\in \mathcal{R}_u}\frac{1}{|R|}\sum_{P\in R}\frac{\#_{P,T}}{|\bpath{P}|}
\]}
The above probability distribution assigns equal probability to routes starting from $u$,  then it assigns equal probability to 
every bundle path of the same route  and, finally, it assigns equal probability to every trace occurrence of the same bundle path.
 

\noindent
The ''stationary'' formulas for general route systems     are     given in Appendix \ref{apx::modular}.
We here give the simpler formulas for balanced route systems.
A Route System $\mathfrak{R} = (\mathcal{B}, \mathcal{R})$  is \emph{balanced} if, for every $u\in \mathcal{S}$,
it holds that \\
$ \left|\{R\in \mathcal{R}| R_{start}=u\}\right|\;=\; \left|\{R\in \mathcal{R}| R_{end}=u\}\right|
$.
  
  \noindent
  We are now able to derive the explicit formulas for the spatial and the destination distributions; observe that such formulas can be computed by counting arguments or by computer    calculations.
  
\begin{prop}\label{balanced_lego_formulas}
Let $\mathfrak{R} = (\mathcal{B}, \mathcal{R})$ be a balanced Route System such that the associated MTM 
$\mathcal{D}[\mathfrak{R}] = (\btripset{\mathfrak{R}}, \Psi[\mathfrak{R}])$ is strongly connected. Then,
$(i)$
The stationary  spatial distribution $\mathfrak{s}$ of $\mathcal{D}[\mathfrak{R}]$ is, for every $u\in \mathcal{S}$,
{\small \[
\mathfrak{s}(u) \;=\; \frac{1}{\Lambda_{\mathrm{b}}[\mathfrak{R}]} \sum_{B\in \mathcal{B}}
\frac{\#_{B, u}}{|B|}\sum_{R\in\mathcal{R}}\frac{\#_{R,B}}{|R|}
 \; \mbox{ with } \; 
 \Lambda_{\mathrm{b}}[\mathfrak{R}] \;=\; \sum_{B\in \mathcal{B}}
\frac{\#B}{|B|} 
\sum_{R\in\mathcal{R}}\frac{\#_{R,B}}{|R|}\\
\]}
$(ii)$
the stationary  destination   distributions $\mathfrak{d}$ of $\mathcal{D}[\mathfrak{R}]$ are, for every $u, v\in \mathcal{S}$,
{\small \[
\mathfrak{d}_u(v) \;=\; \frac{1}{\mathfrak{s}(u)\Lambda_{\mathrm{b}}[\mathfrak{R}]}  \sum_{B\in \mathcal{B}}
\frac{\#_{B, u}}{|B|}\sum_{R\in\mathcal{R}\wedge R_{end} = v}\frac{\#_{R,B}}{|R|}
\]}
\end{prop}
 
 \begin{obs} \label{obs::slowness}
In the formulas of the stationary distributions stated in Prop. \ref{balanced_lego_formulas},  
 the factor $\frac{\#_{B, u}}{|B|}$ is the only one depending on trace segments.  When points are 
 homogeneous square cells of size length $d>0$,  the following important interpretation hold. Given a trace segment 
 $T$ and a cell $u$ of $T$, the ratio  $v_{T,u} = d / (\#_{T, u} \tau)$ (where $\tau$ is the time unit) can be interpreted as the
\emph{(agent)  instanteneous speed} at cell $u$ in trace $T$.  We can thus define the \emph{ instantaneous slowness} as
$ \slow{T,u} =  1/ v_{T,u}$ $ =   (\#_{T, u} \tau)/ d$.  We thus get
$(\#_{B, u})/{|B|} =   ( d / \tau)   \cdot  (\sum_{T \in B} \slow{T,u})/{|B|}$.
Since $d/\tau$ is a constant,  the factor $\#_{B, u} / |B|$ is proportional to the average of the instantaneous slowness in $u$
of bundle $B$: this average will be denoted as $\slow{B,u}$.
 In order to compute the stationary distributions in Prop. \ref{balanced_lego_formulas} of a given 
balanced Route System, we can thus  provide the values of $\slow{B,u}$ rather than specifying all the trace segments.
\end{obs}

\subsection{Application: The DownTown Model} \label{ssec::down}
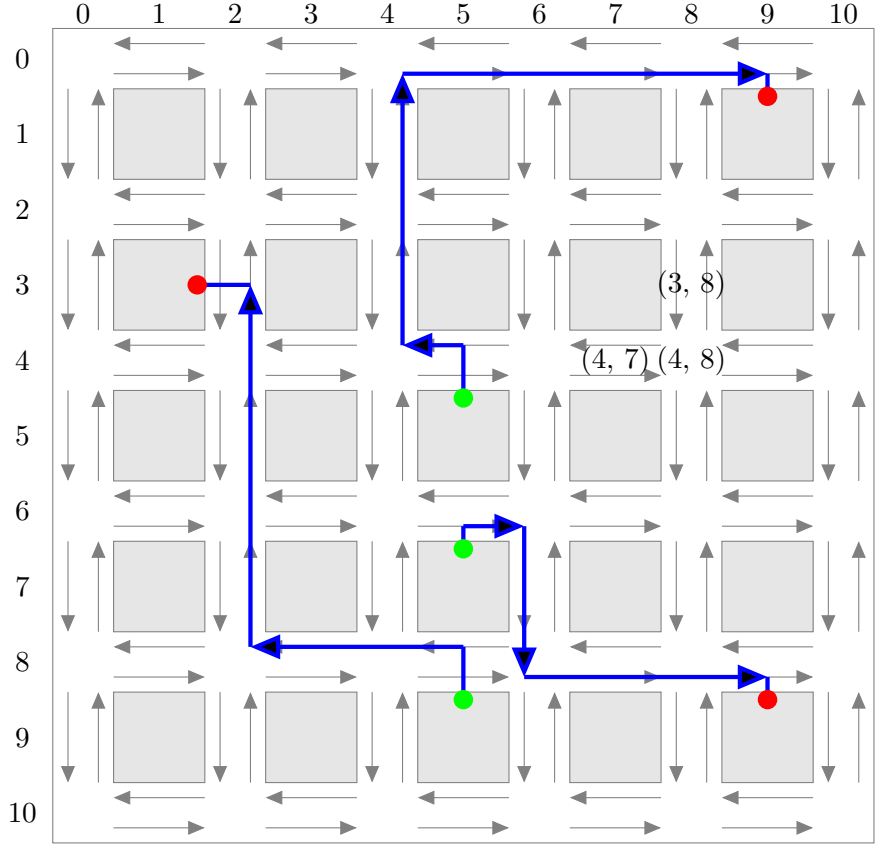
\begin{SCfigure}
\begin{tikzpicture}
[scale=1]
\def\oddlist{0.8,2.8,4.8,6.8,8.8,10.8}
\def\oddlistB{1.2,3.2,5.2,7.2,9.2,11.2}
\def\blocklist{1.4/2.6,3.4/4.6,5.4/6.6,7.4/8.6,9.4/10.6}
\foreach \x / \xx in \blocklist {
    \foreach \y / \yy in \blocklist {
        \draw[gray,fill=black!10] (\x, \y) rectangle (\xx, \yy);
    }
}
\draw[gray] (0.6,0.6) rectangle (11.4,11.4);
\foreach \p / \x in {1/0,2/1,3/2,4/3,5/4,6/5,7/6,8/7,9/8,10/9,11/10} {
    \draw (\p, 11.6) node {\x};
}
\foreach \p / \y in {1/10,2/9,3/8,4/7,5/6,6/5,7/4,8/3,9/2,10/1,11/0} {
    \draw (0.2, \p) node {\y};
}
\def\plist{1.4/2.6,3.4/4.6,5.4/6.6,7.4/8.6,9.4/10.6}
\begin{scope}[gray]
\foreach \x / \xx in \plist {
    \foreach \y in \oddlist {
        \draw[-triangle 45] (\x, \y) -- (\xx, \y);
    }
    \foreach \y in \oddlistB {
        \draw[-triangle 45] (\xx, \y) -- (\x, \y);
    }
}
\foreach \y / \yy in \plist {
    \foreach \x in \oddlist {
        \draw[-triangle 45] (\x, \yy) -- (\x, \y);
    }
    \foreach \x in \oddlistB {
        \draw[-triangle 45] (\x, \y) -- (\x, \yy);
    }
}
\end{scope}
\draw (8, 7) node {(4, 7)};
\draw (9, 8) node {(3, 8)};
\draw (9, 7) node {(4, 8)};
\begin{scope}[ultra thick,draw=blue]
\draw (6, 4.5) -- (6, 4.8);
\draw[draw=green,fill=green] (6, 4.5) circle (1mm);
\draw[-triangle 45] (6, 4.8) -- (6.8, 4.8);
\draw[-triangle 45] (6.8, 4.8) -- (6.8, 2.8);
\draw[-triangle 45] (6.8, 2.8) -- (10, 2.8);
\draw (10, 2.8) -- (10, 2.5);
\draw[draw=red,fill=red] (10, 2.5) circle (1mm);
\end{scope}
\begin{scope}[ultra thick,draw=blue]
\draw (6, 2.5) -- (6, 3.2);
\draw[draw=green,fill=green] (6, 2.5) circle (1mm);
\draw[-triangle 45] (6, 3.2) -- (3.2, 3.2);
\draw[-triangle 45] (3.2, 3.2) -- (3.2, 8);
\draw (3.2, 8) -- (2.5, 8);
\draw[draw=red,fill=red] (2.5, 8) circle (1mm);
\end{scope}
\begin{scope}[ultra thick,draw=blue]
\draw (6, 6.5) -- (6, 7.2);
\draw[draw=green,fill=green] (6, 6.5) circle (1mm);
\draw[-triangle 45] (6, 7.2) -- (5.2, 7.2);
\draw[-triangle 45] (5.2, 7.2) -- (5.2, 10.8);
\draw[-triangle 45] (5.2, 10.8) -- (10, 10.8);
\draw (10, 10.8) -- (10, 10.5);
\draw[draw=red,fill=red] (10, 10.5) circle (1mm);
\end{scope}
\end{tikzpicture}
\caption{The DownTown model with $n = 10$. Directions for each block are shown in gray. 
In blue are shown three possible routes. The starting
cells are in green and the ending cells are in red.}\label{fig_downtown}
\end{SCfigure}

\begin{SCfigure}
\begin{tikzpicture}
[scale=3]
\def\blocklist{1.4/2.6,3.4/4.6}
\foreach \x / \xx in \blocklist {
    \foreach \y / \yy in \blocklist {
        \draw[gray,fill=black!10] (\x, \y) rectangle (\xx, \yy);
    }
}
\draw (1.2, 2) node {$i + 1$};
\draw (1.2, 3) node {$i$};
\draw (1.2, 4) node {$i - 1$};
\draw (2, 4.7) node {$j - 1$};
\draw (3, 4.7) node {$j$};
\draw (4, 4.7) node {$j + 1$};

\draw[-triangle 45,draw=black!20,fill=black!20,line width=1mm] (2.9, 4.6) -- (2.9, 3.4);
\foreach \x in {2.7,2.9} {
    \foreach \y / \n in {3.5/6,3.7/5,3.9/4,4.1/3,4.3/2,4.5/1} {
        \draw[gray] (\x, \y) +(-.1, -.1) rectangle ++(.1, .1);
        \draw (\x, \y) node{\n};
    }
}
\draw[-triangle 45,draw=black!20,fill=black!20,line width=1mm] (3.1, 3.4) -- (3.1, 4.6);
\foreach \x in {3.1,3.3} {
    \foreach \y / \n in {3.5/1,3.7/2,3.9/3,4.1/4,4.3/5,4.5/6} {
        \draw[gray] (\x, \y) +(-.1, -.1) rectangle ++(.1, .1);
        \draw (\x, \y) node{\n};
    }
}

\def\lightred{red!50}
\def\lightgreen{green!40}
\def\lightblue{blue!40}
\def\medblue{blue!80}
\def \lightviolet{violet!40}
\def\medviolet{violet!80}
\def\lightorange{orange!50}

\foreach \x in {3.1,3.3} {
\draw[gray] (\x, 2.5) +(-.1, -.1) rectangle ++(.1, .1);
\draw[gray] (\x, 2.3) +(-.1, -.1) rectangle ++(.1, .1);
\draw[gray] (\x, 2.1) +(-.1, -.1) rectangle ++(.1, .1);
\draw[gray,fill=\lightgreen] (\x, 1.9) +(-.1, -.1) rectangle ++(.1, .1);
\draw[gray] (\x, 1.7) +(-.1, -.1) rectangle ++(.1, .1);
\draw[gray] (\x, 1.5) +(-.1, -.1) rectangle ++(.1, .1);
}
\draw[gray,fill=\lightred] (2.9, 2.5) +(-.1, -.1) rectangle ++(.1, .1);
\draw[gray,fill=\lightred] (2.9, 2.3) +(-.1, -.1) rectangle ++(.1, .1);
\draw[gray,fill=\lightred] (2.9, 2.1) +(-.1, -.1) rectangle ++(.1, .1);
\draw[gray,fill=\lightgreen] (2.9, 1.9) +(-.1, -.1) rectangle ++(.1, .1);
\draw[gray,fill=\lightgreen] (2.9, 1.7) +(-.1, -.1) rectangle ++(.1, .1);
\draw[gray,fill=\lightgreen] (2.9, 1.5) +(-.1, -.1) rectangle ++(.1, .1);

\draw[gray] (2.7, 2.5) +(-.1, -.1) rectangle ++(.1, .1);
\draw[gray] (2.7, 2.3) +(-.1, -.1) rectangle ++(.1, .1);
\draw[gray,fill=\lightred] (2.7, 2.1) +(-.1, -.1) rectangle ++(.1, .1);
\draw[gray] (2.7, 1.9) +(-.1, -.1) rectangle ++(.1, .1);
\draw[gray] (2.7, 1.7) +(-.1, -.1) rectangle ++(.1, .1);
\draw[gray] (2.7, 1.5) +(-.1, -.1) rectangle ++(.1, .1);

\draw[green,line width=1mm] (3.4, 1.9) -- (2.9, 1.9);
\draw[-triangle 45,green,line width=1mm] (2.9, 1.9) -- (2.9, 1.4);
\draw[red,line width=1mm] (2.9, 2.6) -- (2.9, 2.1);
\draw[-triangle 45,red,line width=1mm] (2.9, 2.1) -- (2.6, 2.1);

\draw(2.2, 2.1) node{$B_{E, 3}^{++}(i+1, j)$};
\draw(3.8, 1.9) node{$B_{S, 3}^{-+}(i+1, j)$};

\foreach \x in {2.7,2.9} {
    \foreach \y / \n in {1.5/6,1.7/5,1.9/4,2.1/3,2.3/2,2.5/1} {
        \draw (\x, \y) node{\n};
    }
}
\foreach \x in {3.1,3.3} {
    \foreach \y / \n in {1.5/1,1.7/2,1.9/3,2.1/4,2.3/5,2.5/6} {
        \draw (\x, \y) node{\n};
    }
}

\draw[gray] (3.5, 3.3) +(-.1, -.1) rectangle ++(.1, .1);
\draw[gray] (3.7, 3.3) +(-.1, -.1) rectangle ++(.1, .1);
\draw[gray,fill= \lightgreen] (3.9, 3.3) +(-.1, -.1) rectangle ++(.1, .1);
\draw[gray] (4.1, 3.3) +(-.1, -.1) rectangle ++(.1, .1);
\draw[gray] (4.3, 3.3) +(-.1, -.1) rectangle ++(.1, .1);
\draw[gray] (4.5, 3.3) +(-.1, -.1) rectangle ++(.1, .1);

\draw[gray,fill= \lightgreen] (3.5, 3.1) +(-.1, -.1) rectangle ++(.1, .1);
\draw[gray,fill= \lightgreen] (3.7, 3.1) +(-.1, -.1) rectangle ++(.1, .1);
\draw[gray,fill= \lightgreen] (3.9, 3.1) +(-.1, -.1) rectangle ++(.1, .1);
\draw[gray,fill= \lightred] (4.1, 3.1) +(-.1, -.1) rectangle ++(.1, .1);
\draw[gray,fill= \lightred] (4.3, 3.1) +(-.1, -.1) rectangle ++(.1, .1);
\draw[gray,fill= \lightred] (4.5, 3.1) +(-.1, -.1) rectangle ++(.1, .1);

\draw[gray] (3.5, 2.9) +(-.1, -.1) rectangle ++(.1, .1);
\draw[gray] (3.7, 2.9) +(-.1, -.1) rectangle ++(.1, .1);
\draw[gray] (3.9, 2.9) +(-.1, -.1) rectangle ++(.1, .1);
\draw[gray, fill=\lightred] (4.1, 2.9) +(-.1, -.1) rectangle ++(.1, .1);
\draw[gray] (4.3, 2.9) +(-.1, -.1) rectangle ++(.1, .1);
\draw[gray] (4.5, 2.9) +(-.1, -.1) rectangle ++(.1, .1);

\draw[gray] (3.5, 2.7) +(-.1, -.1) rectangle ++(.1, .1);
\draw[gray] (3.7, 2.7) +(-.1, -.1) rectangle ++(.1, .1);
\draw[gray] (3.9, 2.7) +(-.1, -.1) rectangle ++(.1, .1);
\draw[gray, fill=\lightred] (4.1, 2.7) +(-.1, -.1) rectangle ++(.1, .1);
\draw[gray] (4.3, 2.7) +(-.1, -.1) rectangle ++(.1, .1);
\draw[gray] (4.5, 2.7) +(-.1, -.1) rectangle ++(.1, .1);

\draw[green,line width=1mm] (3.9, 3.4) -- (3.9, 3.1);
\draw[-triangle 45,green,line width=1mm] (3.9, 3.1) -- (3.4, 3.1);
\draw[red,line width=1mm] (4.1, 3.1) -- (4.6, 3.1);
\draw[-triangle 45,red,line width=1mm] (4.1, 3.1) -- (4.1,2.6);

\foreach \x / \n in {3.5/6,3.7/5,3.9/4,4.1/3,4.3/2,4.5/1} {
    \foreach \y in {3.3,3.1} {
        \draw (\x, \y) node{\n};
    }
}
\foreach \x / \n in {3.5/1,3.7/2,3.9/3,4.1/4,4.3/5,4.5/6} {
    \foreach \y in {2.9,2.7} {
        \draw (\x, \y) node{\n};
    }
}

\draw(3.9, 3.5) node{$B_{S, 4}^{--}(i, j+1)$};
\draw(4.1, 2.5) node{$B_{E, 4}^{+-}(i, j+1)$};

\foreach \x in {1.5,1.7,1.9,2.1,2.3,2.5} {
    \draw[gray] (\x, 3.3) +(-.1, -.1) rectangle ++(.1, .1);
    \draw[gray,fill=\lightblue] (\x, 3.1) +(-.1, -.1) rectangle ++(.1, .1);
    \draw[gray,fill=\lightblue] (\x, 2.9) +(-.1, -.1) rectangle ++(.1, .1);
    \draw[gray] (\x, 2.7) +(-.1, -.1) rectangle ++(.1, .1);
}

\draw[-triangle 45,\medblue,line width=1mm] (2.6, 3.1) -- (1.4, 3.1);
\draw[-triangle 45,\medblue,line width=1mm]  (1.4, 2.9) -- (2.6, 2.9);

\foreach \x / \n in {1.5/6,1.7/5,1.9/4,2.1/3,2.3/2,2.5/1} {
    \foreach \y in {3.3,3.1} {
        \draw (\x, \y) node{\n};
    }
}
\foreach \x / \n in {1.5/1,1.7/2,1.9/3,2.1/4,2.3/5,2.5/6} {
    \foreach \y in {2.9,2.7} {
        \draw (\x, \y) node{\n};
    }
}

\draw(2, 3.5) node{$B_{T}^{-}(i, j-1)$};
\draw(2, 2.5) node{$B_{T}^{+}(i, j-1)$};

\draw[gray,fill=black!10] (2.7, 3.3) +(-.1, -.1) rectangle ++(.1, .1);
\draw[gray,fill=black!10] (3.3, 3.3) +(-.1, -.1) rectangle ++(.1, .1);
\draw[gray,fill=black!10] (2.7, 2.7) +(-.1, -.1) rectangle ++(.1, .1);
\draw[gray,fill=black!10] (3.3, 2.7) +(-.1, -.1) rectangle ++(.1, .1);
\draw[gray] (2.9, 3.3) +(-.1, -.1) rectangle ++(.1, .1);
\draw[gray] (3.1, 3.3) +(-.1, -.1) rectangle ++(.1, .1);
\draw[gray,fill=\lightviolet] (2.7, 3.1) +(-.1, -.1) rectangle ++(.1, .1);
\draw[gray,fill= \lightviolet] (3.3, 3.1) +(-.1, -.1) rectangle ++(.1, .1);
\draw[gray,fill= \lightviolet] (2.9,3.1) +(-.1, -.1) rectangle ++(.1, .1);
\draw[gray,fill= \lightviolet] (3.1,3.1) +(-.1, -.1) rectangle ++(.1, .1);
\draw[gray,fill=\lightorange] (2.7, 2.9) +(-.1, -.1) rectangle ++(.1, .1);
\draw[gray,fill=\lightorange] (2.9, 2.9) +(-.1, -.1) rectangle ++(.1, .1);
\draw[gray,fill=\lightorange] (2.9, 2.7) +(-.1, -.1) rectangle ++(.1, .1);
\draw[gray] (3.1, 2.7) +(-.1, -.1) rectangle ++(.1, .1);
\draw[gray] (3.3, 2.9) +(-.1, -.1) rectangle ++(.1, .1);
\draw[gray] (3.1, 2.9) +(-.1, -.1) rectangle ++(.1, .1);

\draw[-triangle 45,\medviolet,line width=1mm] (3.4, 3.1) -- (2.6, 3.1);
\draw[orange,line width=1mm] (2.6, 2.9) -- (2.9, 2.9);
\draw[-triangle 45,orange,line width=1mm] (2.9, 2.9) -- (2.9, 2.6);

\end{tikzpicture}
\caption{
The cross way at $(i, j)$ and its four adjacent blocks are shown, with $m = 6$. The shadows 
of two transit bundles  are shown  blue. The shadows of start bundles,  $--$ and $-+$, are shown in green, the other
two cases (i.e., $++$, and $+-$) are symmetric.
The shadows of end bundles,  $++$ and $+-$, are shown in red,
the other two cases are symmetric. The shadow of $B_{C}^{H,-}(i, j)$ is shown in violet, the other
three straight cross bundles are symmetric.   The shadow of $B_{C}^{H,++}(i, j)$ is 
shown in orange, the other
seven turn cross bundles are symmetric.
}
\label{fig_shadows}
\end{SCfigure}
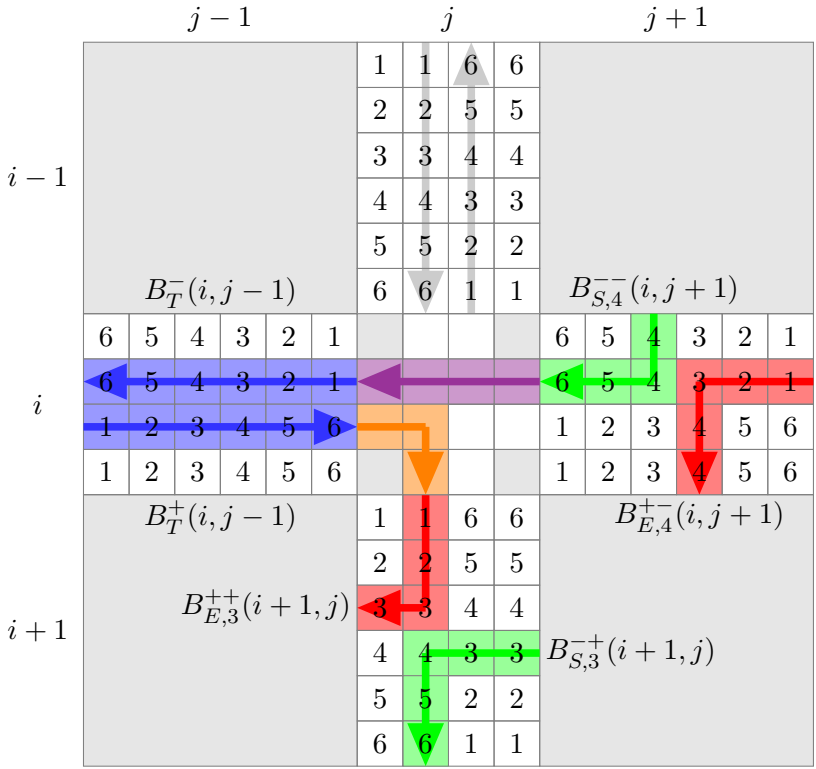

We now use  the Modular Trace Model to describe vehicles that move over a squared \emph{city-like}  support.
This support consists of  a square of $(n+1) \times (n+1)$ crossing \emph{streets} (horizontal and vertical) 
 and \emph{buildings} (where $n$ is an even number).
Buildings are interdicted zones,  while  veichles move and park on the  streets.  Streets
are in turn formed by \emph{parking}   and \emph{transit} cells and 
every transit cells of a given street has its own   direction  (see Fig.s  \ref{fig_downtown} and \ref{fig_shadows}).
  Moreover, a parking cell has its natural direction given by the direction of its closest transit cells.

\noindent
A vehicle   (agent)   moves from one parking cell (the \emph{start}) to another parking one (the \emph{destination})
by choosing at random one of the  feasible paths. To every feasible path,  a set of traces is uniquely
associated that models the different ways    a vehicle may  run over that path: this will also allow to  simulate  
traffic lights on the cross-ways.

\noindent
Every street is an alternating sequence of   \emph{cross-ways} and \emph{blocks}. We enumerate
horizontal streets by an increasing even index $\{0, 2, \ldots , n\}$, starting from the left-top corner.
We do the same for vertical streets as well. In this way, every cross-way gets a pair of coordinates $(i,j)$.
We say that   a direction is \emph{positive} over a horizontal street if it goes from left to right and  while 
the opposite direction is said to be \emph{negative}.  As for vertical streets, the positive direction is the
one going from top to bottom and the opposite is said to be negative  (see Fig.s  \ref{fig_downtown} and \ref{fig_shadows}).

\noindent
Then, the blocks of a horizontal street $i$ will be indexed  from left to right with coordinates $(i,1),
(i,3), (i,5), \ldots $. Similarly,  the block of a vertical street $j$ will be indexed from top to bottom
with coordinates $(1,j), (3,j), \ldots)$.

\noindent

\noindent

\noindent
We now formally introduce the \emph{DownTown Route System} 
$\mathfrak{R}^D = \langle \mathcal B^D, \mathcal R^D \rangle$; let's start with  the bundle set
$\mathcal B^D$.  

\smallskip
\noindent
\textbf{[Blocks.] }   Each (street) block is formed by $4$ stripes of $m$ cells  each 
with indexing shown in Fig. \ref{fig_shadows}.   Two stripes are
for  transit while the two external ones are for parking use. The parking stripe adjacent to
the transit stripe with positive direction is said   \emph{positive parking stripe} while the other
one is said  \emph{negative parking stripe}. \\
For every 
$0 \leqslant i,j \leqslant n$ such that $( i \ \mbox{ odd }  \wedge   j   \mbox{ even } )   \vee
(i  \mbox{  even }  \wedge   j  \mbox{ odd })$, 
Block $(i,j)$ has the following bundles.
         Bundle $\btrans{+}$ whose shadow is the stripe having positive direction;
                              Bundle $\btrans{-}$  is symmetric to $\btrans{+}$ for the negative direction;
                   For each parking cell of index $k$, there are   four start-Bundles $\bstop{S}{++}$,
                     $\bstop{S}{+-}$, $\bstop{S}{--}$,  and $\bstop{S}{-+}$;
                      For each parking cell of index $k$, there are   four end-Bundles $\bstop{E}{++}$,
                     $\bstop{E}{+-}$, $\bstop{E}{--}$,  and $\bstop{E}{-+}$. The shadows of the above bundles 
                      are  shown in Fig.  \ref{fig_shadows}.
              
\noindent
As for the trace segments, thanks to Obs. \ref{obs::slowness}, we only describe the average slowness of the bundles.
For the sake of simplicity, we assume the latter depends only on the cell positions.  For any transit cell of index $k$, the average
slowness in that cell is $\slk(k)$, where $\slk( )$ is an arbitrary positive function. 
The $\slk( )$ function is the same for all transit cells of any of the above bundles. Notice that, by choosing a suitable function
$\slk()$, we can  implement  variable  agent speeds    and  simulate traffic lights  at the cross-ways. For instance, assume there is
a  traffic  light  on a cross-way having  two possible states (red and green): the 2 states alternate at regular fixed time (say 
1 minute). Then the slowness of cells along every of the four  adjacent transit blocks should be an increasing  function of the
distance between the cell (determined by index $k$) and the crossway. The exact form of such increasing function
depends on the average traffic   over that cell (notice that the ratio between the highest and the lowest slowness 
in the same transit block might be order of hundreds).
As for the parking cells, we assume that the average slowness is equal to a positive  constant  $\wait$.

\smallskip
\noindent
\textbf{[Cross-Ways.]} A cross-way is formed by 12 cells as shown in Fig.  \ref{fig_shadows}. We have two types of  associated bundles.\\ For   every 
$ 0 \leqslant i,j \leqslant n$   such that  $(  i  \mbox{ even }   \wedge   j  \mbox{ even } )$, 
we have:
  The 4 straight bundles $\bcross{H}{+}$,  $\bcross{H}{-}$, $\bcross{V}{+}$, and $\bcross{V}{-}$;                               
                      The 8 turn 
                     bundles $\bcross{H}{++}$,  $\bcross{H}{+-}$, $\bcross{H}{--}$
                     $\bcross{H}{-+}$; Moreover,  
                     $\bcross{V}{++}$,   $\bcross{V}{+-}$, $\bcross{V}{--}$, and $\bcross{V}{-+}$. The 
                     relative shadows are shown in Fig. \ref{fig_shadows}. Observe that the first sign indicates the sign of the in-direction and the
                     other indicates the out-direction.
We assume that the average slowness of cross-ways cells is set to a positive constant $\crc$.

\smallskip
\noindent
Let us now introduce the set of DownTown routes $\mathcal R^D$ formed by combining 
the bundles described above.  First of all,  every route contains exactly one bundle-path.
So we   can   describe the bundle-path.  For every pair $\langle c , c' \rangle$
 of parking cells not 
belonging to the same block, there is (only) one bundle-path that goes from $c$ to $c'$.
The structure of a bundle-path has a start-bundle for $c$, followed by an alternating sequence
of block and cross-way bundles, and finally an end-bundle for $c'$, i.e.,
$S,C_1,B_2,C_3, \ldots , B_{k-1}, C_k, E $  with  $ k \geqslant 1$.

\noindent
Notice that two consecutve bundles of a bundle-path sequence belong to two adjacent
cross-ways and blocks, and  the  bundles of a bundle-path are all distinct.
Let us describe the bundle-paths whose starting cells belong to horizontal blocks. The  case of vertical blocks is fully
 symmetric.  let $(i,j)$ be the coordinates of  the 
starting horizontal block and let $(k,z)$ be the coordinates of the ending block.

\noindent
\textbf{[Vertical End Block.] } The bundle-path is easily determined by the following 
\emph{driving directions}:  go straight down  horizontal   street $i$ toward cross-way $(i,z)$;
turn to vertical street $z$ till block $(k,z)$.

\noindent
\textbf{[Horiz. End Block.]} This case yields  in turn two subcases. \\
-   Case $j \neq z$.  If $k=i$
then go straight down to  horizontal street $i$ till block $(i,z)$;  if $k \neq i$
then go to cross-way $(i,j+a)$ (where $a=1$ if $i < k$ and $a = -1$ otherwise) 
and turn  to vertical street $j+a$ towards cross-way $(k,j+a )$, then
turn to  horizontal street $k$ towards block $(k,z)$.  \\
- Case $j = z$. The resulting bundle-path depends on whether the starting cell belongs to
  positive or negative parking stripe. Go to cross-way $(i,j+a)$ 
  (where $a=1$ if we are in the positive case  and $a = -1$ otherwise) 
and turn to vertical street $j+a$ towards cross-way $(k,j+a )$, then
turn to  horizontal street $k$ towards block $(k,z)$.

\noindent
Typical examples of the above paths are shown in Fig. \ref{fig_downtown}.

\noindent
Observe   there are some  cross-way bundles on the border that do not belong to any route.
For the sake of convenience, for every bundle $B \in \mathcal B^D$,
we define 

{\small  \[ \sbg \ =  \ \sum_{R\in \mathcal{R}^{D}}\frac{\#_{R,B}}{|R|}  \]}
Since  every route contains exactly one bundle-path and    a bundle occurs at most once in any bundle-path, then
$\sbg$ equals the number of bundle-paths containing bundle $B$.  
In Appendix \ref{apx:boundlecounting},  we    compute $\sbg$ for each kind of bundle in order to  get 
the stationary spatial distribution given by Prop. \ref{balanced_lego_formulas}. Due to lack of space,
below we   give  such formulas only  for a cell in a horizontal transit block. As for parking cell and
cross-way cells we provide explicit  formulas in Appendix \ref{apx:boundlecounting}.

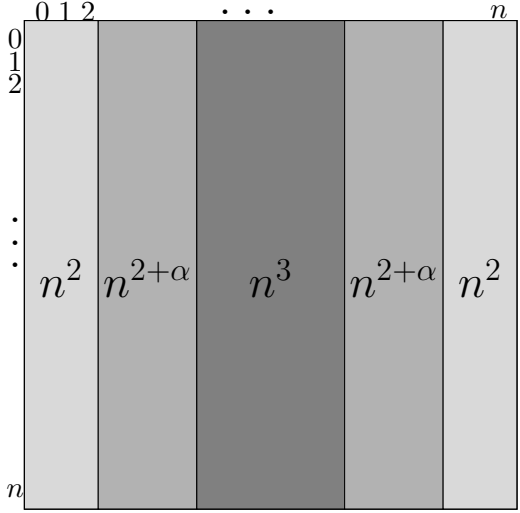
\begin{SCfigure}
\begin{tikzpicture}
[scale=0.6]

\draw (1, 11.6) node {0};
\draw (1.5, 11.6) node {1};
\draw (2, 11.6) node {2};
\draw (5, 11.6) node {\textbf{.}};
\draw (5.5, 11.6) node {\textbf{.}};
\draw (6, 11.6) node {\textbf{.}};
\draw (11, 11.6) node {$n$};
\draw (0.4, 1) node {$n$};
\draw (0.4, 6) node {\textbf{.}};
\draw (0.4, 6.5) node {\textbf{.}};
\draw (0.4, 7) node {\textbf{.}};
\draw (0.4, 10) node {2};
\draw (0.4, 10.5) node {1};
\draw (0.4, 11) node {0};
\def\color{black}
\def\bord{15}
\def\med{20}
\def\center{30}
\def\LL{10.8}
\pgfmathsetmacro{\XB}{0.6 + (\bord/100)*\LL}
\pgfmathsetmacro{\XM}{\XB + (\med/100)*\LL}
\pgfmathsetmacro{\XC}{\XM + (\center/100)*\LL}
\pgfmathsetmacro{\XMM}{\XC + (\med/100)*\LL}
\dw{\color}{15} (0.6,0.6) rectangle (\XB,11.4);
\dw{\color}{30} (\XB,0.6) rectangle (\XM,11.4);
\dw{\color}{50} (\XM,0.6) rectangle (\XC,11.4);
\dw{\color}{30} (\XC,0.6) rectangle (\XMM,11.4);
\dw{\color}{15} (\XMM,0.6) rectangle (11.4,11.4);
\draw[black] (0.6,0.6) rectangle (11.4,11.4);
\pgfmathsetmacro{\XBC}{0.6 + (\XB  - 0.6)/2}
\pgfmathsetmacro{\XMC}{\XB + (\XM  - \XB)/2}
\pgfmathsetmacro{\XCC}{\XM + (\XC  - \XM)/2}
\pgfmathsetmacro{\XMMC}{\XC + (\XMM  - \XC)/2}
\pgfmathsetmacro{\XBBC}{\XMM + (11.4  - \XMM)/2}
\draw (\XBC, 5.7) node {\LARGE $n^2$};
\draw (\XMC, 5.7) node {\LARGE $n^{2 + \alpha}$};
\draw (\XCC, 5.7) node {\LARGE $n^3$};
\draw (\XMMC, 5.7) node {\LARGE $n^{2 + \alpha}$};
\draw (\XBBC, 5.7) node {\LARGE $n^2$};
\end{tikzpicture}
\caption{Asymptotical behaviour of the spatial probability distribution of horizontal positive transit cells ($0 < \alpha < 1$).   The role of  vertical
coordinate $i$ is almost negligible since the only transit direction of such cells is the \emph{positive horizontal} one.  This representation takes no care about the slowness of the cells. Clearly,  in  vertical positive cells, the roles of $i$ and $j$
interchange.
  }\label{fig_dtasymptotic}
\end{SCfigure}

\noindent
Let $\Lambda = \Lambda_{\mathrm{b}}[\mathfrak{R^D}] / m^2$  be the normalization constant
 with $\Lambda_{\mathrm{b}}[\mathfrak{R^D}]$ defined in Prop \ref{balanced_lego_formulas}.  
Let $u$ be a   transit  cell of index  $k$ (with any  $k \in \{1, \ldots , m\}$)
 in the \emph{positive transit stripe} of the horizontal transit  block  $(i,j)$  (i.e. $i$ even and $j$ odd).   
  (with $i \notin \{0,n\}$).
Then 
 {\small 
\[ 
\mathfrak{s}(u)  = \frac{\slk(k)}{\Lambda}\left(a(i,j)+\frac{k}{m}b(j)+\frac{1}{m}c(j)\right) \ \mbox{ where }
\]
\[
a(i,j) = (n-j)(2nj+j+n-i-1)+(n-2)j-\frac{n}{2}+i-2, \]
\[ b(j) =  n(n+1)+\frac{n}{2}-2(n+1)j , \ \mbox{ and } \ 
c(j)  = (n+1)(n+j)+n-3 \]
}
 \noindent
 An informal representation of the asymptotical behaviour of the above function is 
 given in Fig.  \ref{fig_dtasymptotic}.

\noindent

\noindent
\textbf{Acknowledgements.} We are very grateful to Paolo Penna for useful comments.

\newpage
\setcounter{page}{0}
\thispagestyle{empty}

 \appendix
 
 \begin{small}

\section{Proof of Theorem \ref{kernel_stationary}}  \label{apx::kernel_stationary}

In order to prove Theorem  \ref{kernel_stationary}, we need some preliminary lemmas.

\begin{lemma}\label{stationary_lemma}
Let $\mathcal{D} = (\tripset, \Psi)$ be any MTM and let $\pi$ be a stationary distribution of $\mathcal{D}$.
Then, the following properties hold.
\begin{enumerate}[(a)]
\item For every $\langle T, i\rangle\in S(\tripset)$,
\[
\pi(\langle T, i\rangle) \;=\; \pi(\langle T, 1\rangle)
\]
\item For every $T\in \tripset$,
\[
 \pi(\langle T, 1\rangle) \;=\; \psi_{T_{start}}(T)\sum_{T'\in \tripin{T_{start}}} \pi(\langle T', 1\rangle)
\]
\item For every $u\in P(\tripset)$,
\[
\sum_{T\in \tripout{u}} \pi(\langle T, 1\rangle) \;=\; \sum_{T\in \tripin{u}} \pi(\langle T, 1\rangle)
\]
\end{enumerate}
\end{lemma}
\proof
Since $\pi$ is a stationary distribution of $\mathcal{D}$, it holds that, for every $\langle T, i\rangle\in S(\tripset)$,
\[
\pi(\langle T, i\rangle) \;=\; \sum_{\langle T', j\rangle\in S(\tripset)} \pi(\langle T', j\rangle)\Prob{\langle T', j\rangle\rightarrow
\langle T, i\rangle}
\]
If $i > 1$ then $\Prob{\langle T', j\rangle\rightarrow\langle T, i\rangle} = 0$ for any $T'\neq T$ or $j \neq i-1$.
Thus, for every $1 < i \leqslant |T| - 1$,
\[
\pi(\langle T, i\rangle) \;=\;  \pi(\langle T, i-1\rangle)
\]
This implies that, for every $\langle T, i\rangle\in S(\tripset)$,
\[
\pi(\langle T, i\rangle) = \pi(\langle T, 1\rangle)
\]
and property (a) is proved.

If instead $i = 1$ then $\Prob{\langle T', j\rangle\rightarrow\langle T, 1\rangle} = 0$ whenever $T'_{end}
\neq T_{start}$ or $j \neq |T'| - 1$. Thus,
\begin{eqnarray*}
\pi(\langle T, 1\rangle)
& = &
 \sum_{T'\in \tripin{T_{start}}} \pi(\langle T', |T'| - 1\rangle)\psi_{T_{start}}(T) \\
 & = &
\psi_{T_{start}}(T)\sum_{T'\in \tripin{T_{start}}} \pi(\langle T', 1\rangle) \quad \text{(from property (a))}
\end{eqnarray*}
This proves property (b).

From property (b), it derives that
\begin{eqnarray*}
\sum_{T\in \tripout{u}} \pi(\langle T, 1\rangle)
&= &
\sum_{T\in \tripout{u}} \sum_{T'\in \tripin{u}} \pi(\langle T', |T'| - 1\rangle)\psi_{u}(T)\\
& = &
\sum_{T\in \tripout{u}} \psi_{u}(T) \sum_{T'\in \tripin{u}} \pi(\langle T', |T'| - 1\rangle)\\
& = &
\sum_{T'\in \tripin{u}} \pi(\langle T', |T'| - 1\rangle)\\
& = &
\sum_{T'\in \tripin{T_{start}}} \pi(\langle T', 1\rangle)
\end{eqnarray*}
and property (c) is proved.
\qed

\begin{lemma}\label{sigma2pi_lemma}
Let $\mathcal{D} = (\tripset, \Psi)$ be any MTM and let $\sigma$ be a stationary distribution of $\kernel{\mathcal{D}}$.
Let $\pi : S(\tripset) \rightarrow \mathbb{R}$ be defined as follows:
\[
\forall \langle T, i\rangle \in S(\tripset)\qquad
\pi(\langle T, i\rangle) \;=\; \frac{1}{\sum_{u\in P(\tripset)} \sigma(u)\Lambda_{\Psi}(u)}\sigma(T_{start}) \psi_{T_{start}}(T)
\]
Then, $\pi$ is a stationary distribution of $\mathcal{D}$.
\end{lemma}
\proof
First of all, we verify that $\pi$ is a probability distribution over $S(\tripset)$:
\begin{eqnarray*}
\sum_{\langle T, i\rangle \in S(\tripset)} \pi(\langle T, i\rangle)
& = &
\sum_{\langle T, i\rangle \in S(\tripset)}  \frac{1}{\sum_{u\in P(\tripset)} \sigma(u)\Lambda_{\Psi}(u)}\sigma(T_{start})
 \psi_{T_{start}}(T)\\
& = &
 \frac{1}{\sum_{u\in P(\tripset)} \sigma(u)\Lambda_{\Psi}(u)} \sum_{\langle T, i\rangle \in S(\tripset)} \sigma(T_{start})
 \psi_{T_{start}}(T)\\
 & = &
  \frac{1}{\sum_{u\in P(\tripset)} \sigma(u)\Lambda_{\Psi}(u)} \sum_{u\in P(\tripset)}\sum_{T\in\tripout{u}}\sum_{i = 1}^{|T| - 1}
  \sigma(u)\psi_{u}(T)\\
  & = &
   \frac{1}{\sum_{u\in P(\tripset)} \sigma(u)\Lambda_{\Psi}(u)} \sum_{u\in P(\tripset)} \sigma(u)
   \sum_{T\in\tripout{u}}(|T| - 1) \psi_{u}(T)\\
   & = &
   \frac{1}{\sum_{u\in P(\tripset)} \sigma(u)\Lambda_{\Psi}(u)} \sum_{u\in P(\tripset)} \sigma(u)\Lambda_{\Psi}(u)\\
   & = & 1
\end{eqnarray*}
For proving that $\pi$ is a stationary distribution of $\mathcal{D}$
it suffices to show that, for every $s \in S(\tripset)$,
\[
\pi(s) = \sum_{r \in S(\tripset)} \pi(r)\Prob{r \rightarrow s}
\]
For every $s\in S(\tripset)$, let $\lambda(s) = \sum_{r \in S(\tripset)} \pi(r)\Prob{r \rightarrow s}$.
Let $\langle T, i\rangle$ be any state in $S(\tripset)$, we distinguish two cases.
\begin{description}
\item[$(1 < i \leqslant |T| - 1)$:] In this case $\Prob{\langle T', j\rangle \rightarrow \langle T, i\rangle} = 0$ for any $T'\neq T$ and
$\Prob{\langle T, j\rangle \rightarrow \langle T, i\rangle} = 0$ for any $j \neq i - 1$.
Thus,
\begin{eqnarray*}
\lambda(\langle T, i\rangle)
& = &
\sum_{T'\in \tripset}\sum_{j = 1}^{|T'| - 1}\pi(\langle T', j\rangle)\Prob{\langle T', j\rangle \rightarrow \langle T, i\rangle} \\
 & = &
 \sum_{j = 1}^{|T| - 1}\pi(\langle T, j\rangle)\Prob{\langle T, j\rangle \rightarrow \langle T, i\rangle}\\
 & = &
 \pi(\langle T, i-1\rangle)\Prob{\langle T, i-1\rangle \rightarrow \langle T, i\rangle}\\
 & = &
 \pi(\langle T, i-1\rangle) \;=\; \pi(\langle T, i\rangle)
\end{eqnarray*}
where the last equality derives from the definition of $\pi$.
\item[$(i = 1)$:] In this case $\Prob{\langle T', j\rangle \rightarrow \langle T, 1\rangle} = 0$ for any $T'$ and $j$ such that
$T'_{end} \neq T_{start}$ or $j \neq |T'| - 1$. Thus,
\begin{eqnarray*}
\lambda(\langle T, 1\rangle)
& = &
\sum_{T'\in \tripset}\sum_{j = 1}^{|T'| - 1}\pi(\langle T', j\rangle)\Prob{\langle T', j\rangle \rightarrow \langle T, 1\rangle} \\
& = &
\sum_{T'\in\tripin{T_{start}}}\pi(\langle T', |T'| - 1\rangle)\Prob{\langle T', |T'| - 1\rangle \rightarrow \langle T, 1\rangle} \\
& = &
\sum_{T'\in\tripin{T_{start}}}\pi(\langle T', |T'| - 1\rangle)\psi_{T_{start}}(T)\\
& = &
\psi_{T_{start}}(T)\sum_{T'\in \tripin{T_{start}}}\pi(\langle T', |T'| - 1\rangle)\\
& = &
 \psi_{T_{start}}(T)\sum_{T'\in \tripin{T_{start}}}
\frac{1}{\sum_{u\in P(\tripset)} \sigma(u)\Lambda_{\Psi}(u)}\sigma(T'_{start}) \psi_{T'_{start}}(T') \\
& = &
\frac{\psi_{T_{start}}(T)}{\sum_{u\in P(\tripset)} \sigma(u)\Lambda_{\Psi}(u)}\sum_{u\in P(\tripset)}\sum_{T'\in \trips{u}{T_{start}}} \sigma(u)\psi_{u}(T') \\
& = &
 \frac{\psi_{T_{start}}(T)}{\sum_{u\in P(\tripset)} \sigma(u)\Lambda_{\Psi}(u)}
\sum_{u\in P(\tripset)}\sigma(u)\sum_{T'\in \trips{u}{T_{start}}} \psi_{u}(T')
\end{eqnarray*}
Observe that $\sum_{T'\in \trips{u}{T_{start}}} \psi_{u}(T')$ is equals to $\Prob{u\rightarrow T_{start}}$ (i.e., the transition
probability from state $u$ to state $T_{start}$ of the Markov chain $\kernel{\mathcal{D}}$). It follows that
\[
 \lambda(\langle T, 1\rangle)  \;=\; \frac{\psi_{T_{start}}(T)}{\sum_{u\in P(\tripset)} \sigma(u)\Lambda_{\Psi}(u)}
 \sum_{u\in P(\tripset)}\sigma(u)\Prob{u\rightarrow T_{start}}
\]
Since $\sigma$ is a stationary probability distribution of $\kernel{\mathcal{D}}$, it holds that
\[
\sum_{u\in P(\tripset)}\sigma(u)\Prob{u\rightarrow T_{start}} \;=\; \sigma(T_{start})
\]
Hence,
\[
\lambda(\langle T, 1\rangle)  \;=\; \frac{\psi_{T_{start}}(T)}{\sum_{u\in P(\tripset)} \sigma(u)\Lambda_{\Psi}(u)}
\sigma(T_{start}) \;=\; \pi(\langle T, 1\rangle)
\]
\end{description}
\qed

\begin{lemma}\label{pi2sigma_lemma}
Let $\mathcal{D} = (\tripset, \Psi)$ be any MTM and let $\pi$ be a stationary distribution of $\mathcal{D}$.
Let $\sigma : P(\tripset) \rightarrow \mathbb{R}$ be defined as follows:
\[
\forall u\in P(\tripset) \qquad
\sigma(u) \; =\; \frac{1}{\sum_{T\in \tripset} \pi(\langle T, 1\rangle)}\sum_{T\in \tripout{u}} \pi(\langle T, 1\rangle)
\]
Then, $\sigma$ is a stationary distribution of $\kernel{\mathcal{D}}$.
\end{lemma}
\proof
First of all we prove that $\sigma$ is a probability distribution over $P(\tripset)$:
\begin{eqnarray*}
\sum_{u\in P(\tripset)} \sigma(u)
& = &
\sum_{u\in P(\tripset)} \frac{1}{\sum_{T\in \tripset} \pi(\langle T, 1\rangle)}\sum_{T\in \tripout{u}} \pi(\langle T, 1\rangle)\\
& = &
\frac{1}{\sum_{T\in \tripset} \pi(\langle T, 1\rangle)} \sum_{u\in P(\tripset)} \sum_{T\in \tripout{u}} \pi(\langle T, 1\rangle)\\
& = &
\frac{1}{\sum_{T\in \tripset} \pi(\langle T, 1\rangle)} \sum_{T\in \tripset} \pi(\langle T, 1\rangle)\\
& = &
1
\end{eqnarray*}
For proving that $\sigma$ is a stationary distribution of $\kernel{\mathcal{D}}$ it suffices to
show that, for every $u\in P(\tripset)$,
\[
\sigma(u) = \sum_{v \in P(\tripset)} \sigma(v)\Prob{v \rightarrow u}
\]
We use the following abbreviations $\varsigma(u) = \sum_{v \in P(\tripset)} \sigma(v)\Prob{v \rightarrow u}$
and $\alpha = \frac{1}{\sum_{T\in \tripset} \pi(\langle T, 1\rangle)}$.
It holds that
\begin{eqnarray*}
\varsigma(u)
& = &
\sum_{v \in P(\tripset)}  \alpha\sum_{T\in \tripout{v}}
\pi(\langle T, 1\rangle) \Prob{v \rightarrow u}\\
 & = &
\alpha\sum_{v \in P(\tripset)}  \sum_{T\in \tripout{v}}
\pi(\langle T, 1\rangle) \Prob{v \rightarrow u}\\
& = &
\alpha\sum_{v \in P(\tripset)}  \Prob{v \rightarrow u}\sum_{T\in \tripout{v}}
\pi(\langle T, 1\rangle) \\
 & = &
 \alpha\sum_{v \in P(\tripset)} \left( \sum_{T\in \trips{v}{u}}\psi_v(T) \right)
 \sum_{T\in \tripout{v}} \pi(\langle T, 1\rangle) \\
 & = &
 \alpha\sum_{v \in P(\tripset)} \left( \sum_{T\in \trips{v}{u}}\psi_v(T) \right)
\sum_{T'\in \tripin{v}} \pi(\langle T', 1\rangle)  \quad \text{(by Lemma~\ref{stationary_lemma}~(c))}\\
 & = &
  \alpha\sum_{v \in P(\tripset)} \sum_{T\in \trips{v}{u}}
\sum_{T'\in \tripin{v}} \pi(\langle T', 1\rangle)\psi_v(T) \\
 & = &
 \alpha\sum_{v \in P(\tripset)} \sum_{T\in \trips{v}{u}}
\pi(\langle T, 1\rangle) \quad  \text{(by Lemma~\ref{stationary_lemma}~(b))}\\
 & = &
 \alpha \sum_{T\in \tripin{u}}\pi(\langle T, 1\rangle) \\
 & = &
 \alpha \sum_{T\in \tripout{u}}\pi(\langle T, 1\rangle) \quad  \text{(by Lemma~\ref{stationary_lemma}~(c))}\\
 & = &
 \sigma(u)
\end{eqnarray*}
\qed

\noindent
\textbf{Proof of Theorem \ref{kernel_stationary}.}
 Consider property (a). The ``if'' part is equivalent to Lemma~\ref{sigma2pi_lemma}. Now we prove the ``only if'' part.
Let $\pi$ be a stationary distribution of $\mathcal{D}$. From Lemma~\ref{pi2sigma_lemma} the map
$\sigma : P(\tripset)\rightarrow\mathbb{R}$ defined as
\[
\forall u\in P(\tripset) \qquad
\sigma(u) \; =\; \frac{1}{\sum_{T\in \tripset} \pi(\langle T, 1\rangle)}\sum_{T\in \tripout{u}} \pi(\langle T, 1\rangle)
\]
is a stationary distribution of $\kernel{\mathcal{D}}$. For every $\langle T, i\rangle \in S(\tripset)$, let
\[
\lambda(\langle T, i\rangle) \;=\; \frac{1}{\sum_{u\in P(\tripset)} \sigma(u)\Lambda_{\Psi}(u)}\sigma(T_{start}) \psi_{T_{start}}(T)
\]
From this definition and from Lemma~\ref{stationary_lemma}~(a), for showing that $\lambda = \pi$ it suffices to
prove that, for every $T\in\tripset$, $\lambda(\langle T, 1\rangle) = \pi(\langle T, 1\rangle)$. It holds that
\begin{eqnarray}
\lambda(\langle T, 1\rangle)
& = & \nonumber
 \frac{1}{\sum_{u\in P(\tripset)} \sigma(u)\Lambda_{\Psi}(u)}\sigma(T_{start}) \psi_{T_{start}}(T)\\
 & = & \nonumber
 \frac{1}{\sum_{u\in P(\tripset)}\frac{1}{\sum_{T'\in \tripset} \pi(\langle T', 1\rangle)}\sum_{T'\in \tripout{u}} \pi(\langle T', 1\rangle)\Lambda_{\Psi}(u)}\sigma(T_{start}) \psi_{T_{start}}(T)\\
 & = & \label{eq1}
\frac{\sum_{T'\in \tripset} \pi(\langle T', 1\rangle)}{\sum_{u\in P(\tripset)}\sum_{T'\in \tripout{u}} \pi(\langle T', 1\rangle)
\Lambda_{\Psi}(u)}\sigma(T_{start}) \psi_{T_{start}}(T)
\end{eqnarray}
Now, observe that
\begin{eqnarray*}
\sum_{u\in P(\tripset)}\sum_{T'\in \tripout{u}} \pi(\langle T', 1\rangle)\Lambda_{\Psi}(u)
& = &
\sum_{u\in P(\tripset)}\sum_{T'\in \tripout{u}} \pi(\langle T', 1\rangle)\sum_{T''\in \tripout{u}} (|T''| - 1)\psi_u(T'')\\
& = &
\sum_{u\in P(\tripset)}\sum_{T''\in \tripout{u}} (|T''| - 1)\psi_u(T'')\sum_{T'\in \tripout{u}} \pi(\langle T', 1\rangle)\\
\end{eqnarray*}
Since $\pi$ is a stationary distribution of $\mathcal{D}$, from Lemma~\ref{stationary_lemma} (b) and (c),
it holds that (since $T''\in \tripout{u}$)
\[
\psi_u(T'')\sum_{T'\in \tripout{u}} \pi(\langle T', 1\rangle) \;=\; \pi(\langle T'', 1\rangle)
\]
Thus,
\begin{eqnarray*}
\sum_{u\in P(\tripset)}\sum_{T'\in \tripout{u}} \pi(\langle T', 1\rangle)\Lambda_{\Psi}(u)
& = &
\sum_{u\in P(\tripset)}\sum_{T''\in \tripout{u}} (|T''| - 1)\psi_u(T'')\sum_{T'\in \tripout{u}} \pi(\langle T', 1\rangle)\\
& = &
\sum_{u\in P(\tripset)}\sum_{T''\in \tripout{u}} (|T''| - 1)\pi(\langle T'', 1\rangle)\\
& = & \sum_{u\in P(\tripset)}\sum_{T''\in \tripout{u}} \sum_{i = 1}^{|T''| - 1}\pi(\langle T'', i\rangle)
\quad \text{(by Lemma~\ref{stationary_lemma}~(a))}\\
& = &
\sum_{\langle T'', i \rangle\in\tripset} \pi(\langle T'', i\rangle)\\
& = &
1
\end{eqnarray*}
By combining this with Eq.~\ref{eq1}, we have that
\begin{eqnarray*}
\lambda(\langle T, 1\rangle)
& = &
\sum_{T'\in \tripset} \pi(\langle T', 1\rangle)\sigma(T_{start}) \psi_{T_{start}}(T)\\
& = &
\psi_{T_{start}}(T) \sigma(T_{start}) \sum_{T'\in \tripset} \pi(\langle T', 1\rangle)\\
& = &
\psi_{T_{start}}(T)\frac{1}{\sum_{T''\in \tripset} \pi(\langle T'', 1\rangle)}\sum_{T''\in \tripout{T_{start}}} \pi(\langle T'', 1\rangle)
 \sum_{T'\in \tripset} \pi(\langle T', 1\rangle) \\
 & = &
 \psi_{T_{start}}(T)\sum_{T''\in \tripout{T_{start}}} \pi(\langle T'', 1\rangle)\\
 & = &
 \pi(\langle T, 1\rangle) \quad \text{(by Lemma~\ref{stationary_lemma}~(b) and (c))}
\end{eqnarray*}

Now consider property (b). The ``if'' part is equivalent to Lemma~\ref{pi2sigma_lemma}. It remains to prove
the ``only if'' part. Let $\sigma$ be a stationary distribution of $\kernel{\mathcal{D}}$. From Lemma~\ref{sigma2pi_lemma}
the map $\pi : S(\tripset)\rightarrow\mathbb{R}$ defined as
\[
\forall \langle T, i\rangle \in S(\tripset)\qquad
\pi(\langle T, i\rangle) \;=\; \frac{1}{\sum_{u\in P(\tripset)} \sigma(u)\Lambda_{\Psi}(u)}\sigma(T_{start}) \psi_{T_{start}}(T)
\]
is a stationary distribution of $\mathcal{D}$. For every $u\in P(\tripset)$, let
\[
\varsigma(u) \;=\; \frac{1}{\sum_{T\in \tripset} \pi(\langle T, 1\rangle)}\sum_{T\in \tripout{u}} \pi(\langle T, 1\rangle)
\]
We will prove that $\varsigma = \sigma$. First of all, observe that
\begin{eqnarray*}
\sum_{T\in \tripset} \pi(\langle T, 1\rangle)
& = &
\sum_{T\in \tripset} \frac{1}{\sum_{u\in P(\tripset)} \sigma(u)\Lambda_{\Psi}(u)}\sigma(T_{start}) \psi_{T_{start}}(T)\\
& = &
\frac{1}{\sum_{u\in P(\tripset)} \sigma(u)\Lambda_{\Psi}(u)} \sum_{T\in \tripset} \sigma(T_{start}) \psi_{T_{start}}(T)\\
& = &
\frac{1}{\sum_{u\in P(\tripset)} \sigma(u)\Lambda_{\Psi}(u)} \sum_{u\in P(\tripset)}\sum_{T\in \tripout{u}}
\sigma(u) \psi_u(T)\\
& = &
\frac{1}{\sum_{u\in P(\tripset)} \sigma(u)\Lambda_{\Psi}(u)} \sum_{u\in P(\tripset)}\sigma(u)\sum_{T\in \tripout{u}} \psi_u(T)\\
& = &
\frac{1}{\sum_{u\in P(\tripset)} \sigma(u)\Lambda_{\Psi}(u)} \sum_{u\in P(\tripset)}\sigma(u) \qquad \text{(since
$\sum_{T\in \tripout{u}} \psi_u(T) = 1$)}\\
& = &
\frac{1}{\sum_{u\in P(\tripset)} \sigma(u)\Lambda_{\Psi}(u)}  \qquad \text{(since
$\sum_{u\in P(\tripset)}\sigma(u) = 1$)}
\end{eqnarray*}
Thus, we have
\begin{eqnarray*}
\varsigma(u)
& = &
\frac{1}{\sum_{T\in \tripset} \pi(\langle T, 1\rangle)}\sum_{T\in \tripout{u}} \pi(\langle T, 1\rangle) \\
& = &
\sum_{u\in P(\tripset)} \sigma(u)\Lambda_{\Psi}(u)\sum_{T\in \tripout{u}} \pi(\langle T, 1\rangle)\\
& = &
\sum_{u\in P(\tripset)} \sigma(u)\Lambda_{\Psi}(u)\sum_{T\in \tripout{u}}
\frac{1}{\sum_{u\in P(\tripset)} \sigma(u)\Lambda_{\Psi}(u)}\sigma(T_{start}) \psi_{T_{start}}(T)\\
& = &
\sum_{T\in \tripout{u}} \sigma(T_{start}) \psi_{T_{start}}(T)\\
& = &
\sum_{T\in \tripout{u}} \sigma(u) \psi_u(T)\\
& = &
\sigma(u) \sum_{T\in \tripout{u}} \psi_u(T)\\
& = &
\sigma(u) \qquad \text{(since $\sum_{T\in \tripout{u}} \psi_u(T) = 1$)}
\end{eqnarray*}
\qed

\section{Proof of Corollary \ref{kernel_stationary_cor}} \label{apx::kercor} 
 
Since $\kernel{\mathcal{D}}$ is a finite Markov chain, property (a) is a consequence of Theorem~\ref{kernel_stationary}~(a)
and a standard result that says any finite Markov chain has a stationary distribution. Although it is a standard result, it is not so
easy to find a clear reference for it. Thus, we outline a proof of this result.
Let $\mathcal{M} = (S, P)$ be any Markov chain with $|S| < \infty$. Let $n = |S|$ and let $P = (p_{ij})_{i, j\in S}$. Consider
the subset of the $n$-dimensional Euclidean space $U = \{ (x_1,x_2\ldots,x_n)\;|\; \forall i\; x_i \geqslant 0 \wedge
\sum_{i=1}^n x_i = 1\}$. Let $L : U \rightarrow U$ be the map defined as follows:
\[
\forall \overline{x} = (x_1,\ldots,x_n)\quad L(\overline{x}) = (\sum_{j=1}^np_{j1}x_j, \sum_{j=1}^np_{j2}x_j,\ldots,
\sum_{j=1}^np_{jn}x_j)
\]
It is easy to verify that, $\forall \overline{x}\in U$, $L(\overline{x}) \in U$. Indeed, $\sum_{j=1}^np_{ji}x_j \geqslant 0$ for all $i$.
Moreover $\sum_{i=1}^n \sum_{j=1}^np_{ji}x_j =  \sum_{j=1}^n x_j\sum_{i=1}^n p_{ji} = \sum_{j=1}^n x_j = 1$, since
$\sum_{i=1}^n p_{ji} = 1$ for all $i$. Observe that any fixed point of $L$, that is, a point $\overline{u} \in U$ such that
$L(\overline{u}) = \overline{u}$, is a stationary distribution for the Markov chain $\mathcal{M}$. The existence of at least a
fixed point is guaranteed by Brouwer's fixed point theorem (see for instance \cite{W_BF}):
\begin{quote}
Every continuous map from a convex compact subset $K$ of a Euclidean space to $K$ itself has a fixed point.
\end{quote}
Indeed, it is easy to see that $U$ is compact and convex and $L$ is a continuous map.

\noindent
Property (b) immediately derives from Theorem~\ref{kernel_stationary}~(a).

\qed

\section{Proof of Proposition \ref{kernel_general}}  \label{apx::kern-gen}
 
Let $P = (p_{uv})$. Define $\tripset = \{ (u, v) \;|\; p_{uv} > 0\}$. Observe that for every $(u, v) \in \tripset$,
there exists at least a $w\in S$ such that $(v, w) \in \tripset$, since $\sum_{x\in S} p_{vx} = 1$. This implies that
$\tripset$ is endless. Also notice that $P(\tripset) = S$. For each $u\in P(\tripset)$, define $\psi_u(u,v) = p_{uv}$
for every $(u,v)\in \tripout{u}$. Clearly $\psi_u$ is a probability distribution over $\tripout{u}$. Thus,
$\Psi = \{\psi_u\}_{u\in P(\tripset)}$ is a TSR for $\tripset$. As a consequence $\mathcal{D} = (\tripset, \Psi)$
is a MTM. Since for every $u, v\in P(\tripset)$, $\sum_{T\in\trips{u}{v}} \psi_u(T) = p_{uv}$, it holds that
$\kernel{\mathcal{D}} = \mathcal{M}$.

\qed

\section{Proof of Theorem \ref{uniqueness}}

\begin{lemma}\label{irreducible}
If $\mathcal{D}$ is a strongly connected MTM then $\kernel{\mathcal{D}}$ is an irreducible Markov chain.
\end{lemma}

\proof
Recall that a Markov chain is said to be irreducible if it is possible to go
from every state to every state (not necessarily in one step). Consider any two states $u,v\in P(\tripset)$
of $\kernel{\mathcal{D}}$. Since $\mathcal{D}$ is a strongly connected, there exists a
sequence of points of $P(\tripset)$
$(z_0, z_1,\ldots,z_k)$ such that $z_0 = u$, $z_k = v$, and, for every $i = 0,1,\ldots, k-1$,
\[
\Prob{z_i \rightarrow z_{i+1}} \;=\; \sum_{T\in \trips{z_i}{z_i+1}} \psi_{z_i}(T) \;>\; 0
\]
Thus,
\[
\Prob{u\rightarrow v}^{(k)} \;\geqslant\; \prod_{i=0}^{k-1} \Prob{z_i \rightarrow z_{i+1}} \;>\; 0
\]
where $\Prob{u\rightarrow v}^{(k)}$ denotes the probability of going from $u$ to $v$ in $k$ steps.
Thus, there is a positive probability of reaching $v$ from $u$.

\qed

\noindent
\textbf{Proof of Theorem \ref{uniqueness}.}
 A standard result says that a finite irreducible Markov chain has a unique stationary distribution (see for
instance \cite{GS97} Chapter 11 - Theorem 11.10, here irreducible chains are called ergodic chains).
From this result and Lemma~\ref{irreducible}, it follows that $\kernel{\mathcal{D}}$ has a unique
stationary distribution. Then, from Corollary~\ref{kernel_stationary_cor}~(b), also $\mathcal{D}$
has a unique stationary distribution.
\qed

\section{Proof of  Theorem \ref{uniformity} } \label{apx::uniformness}

We first need the following 

\begin{lemma} \label{uniformity_lemma}
Let $\mathcal{D} = (\tripset, \Psi)$ be any MTM that is both uniformly selective and balanced,
then $\kernel{\mathcal{D}}$ has the following stationary distribution $\sigma$:
$\forall u\in P(\tripset)$, 
$\sigma(u) \;=\; \frac{|\tripout{u}|}{|\tripset|}$
\end{lemma}

\proof It holds that,
for every $u\in P(\tripset)$,
\begin{eqnarray*}
\sum_{v\in P(\tripset)} \sigma(v)\Prob{v \rightarrow u}
& = &
\sum_{v\in P(\tripset)} \sigma(v) \sum_{T\in\trips{v}{u}} \psi_v(T) \\
& = &
\sum_{v\in P(\tripset)} \sigma(v) \frac{|\trips{v}{u}|}{|\tripout{v}|} \qquad \text{(since $\mathcal{D}$ is uniformly selective)}\\
& = &
\sum_{v\in P(\tripset)} \frac{|\tripout{v}|}{|\tripset|} \frac{|\trips{v}{u}|}{|\tripout{v}|} \\
& = &
\sum_{v\in P(\tripset)} \frac{|\trips{v}{u}|}{|\tripset|} \\
& = &
\frac{|\tripin{u}|}{|\tripset|} \\
& = &
\frac{|\tripout{u}|}{|\tripset|} \qquad \text{(since $\mathcal{D}$ is balanced)}\\
& = & \sigma(u)
\end{eqnarray*}

\qed

\noindent
\textbf{Proof of Theorem \ref{uniformity}.}
 Let $\pi$ be a uniform stationary distribution of $\mathcal{D}$. Then
\[
\forall \langle T, i\rangle \qquad \pi(\langle T, i \rangle) \;=\; \frac{1}{|S(\tripset)|}
\]
From Lemma~\ref{stationary_lemma}~(c), it immediately derives that $\mathcal{D}$ is balanced.
The uniformly selectiveity derives from Lemma~\ref{stationary_lemma}~(b).

Conversely, assume that $\mathcal{D}$ is both uniformly selective and balanced.
From Lemma~\ref{uniformity_lemma}, the following probability
distribution $\sigma$ over $P(\tripset)$:
\[
\forall u\in P(\tripset)\qquad \sigma(u) \;=\; \frac{|\tripout{u}|}{|\tripset|}
\]
is stationary for $\kernel{\mathcal{D}}$.
Thus, by Lemma~\ref{sigma2pi_lemma}, the following map $\pi$ is a stationary distribution of
$\mathcal{D}$:
\[
\forall \langle T, i\rangle \in S(\tripset)\qquad
\pi(\langle T, i\rangle) \;=\; \frac{1}{\sum_{u\in P(\tripset)} \sigma(u)\Lambda_{\Psi}(u)}\sigma(T_{start}) \psi_{T_{start}}(T)
\]
Firstly, observe that the following holds:
\begin{eqnarray*}
\sum_{u\in P(\tripset)} \sigma(u)\Lambda_{\Psi}(u)
& = &
\sum_{u\in P(\tripset)} \frac{|\tripout{u}|}{|\tripset|}\sum_{T\in \tripout{u}}(|T| - 1)\psi_u(T) \\
& = &
\sum_{u\in P(\tripset)} \frac{|\tripout{u}|}{|\tripset|}\sum_{T\in \tripout{u}}(|T| - 1)\frac{1}{|\tripout{u}|}\\
& = &
\frac{1}{|\tripset|}\sum_{u\in P(\tripset)} \sum_{T\in \tripout{u}}(|T| - 1)\\
& = &
\frac{|S(\tripset)|}{|\tripset|}
\end{eqnarray*}
Then, it holds that, for every $\langle T, i\rangle \in S(\tripset)$,
\begin{eqnarray*}
\pi(\langle T, i\rangle)
& = &
\frac{1}{\sum_{u\in P(\tripset)} \sigma(u)\Lambda_{\Psi}(u)}\sigma(T_{start}) \psi_{T_{start}}(T)\\
& = &
\frac{|\tripset|}{|S(\tripset)|}\sigma(T_{start}) \psi_{T_{start}}(T)\\
& = &
\frac{|\tripset|}{|S(\tripset)|}\frac{|\tripout{T_{start}}|}{|\tripset|} \frac{1}{|\tripout{T_{start}}|}\\
& = &
\frac{1}{|S(\tripset)|}
\end{eqnarray*}
Thus, $\pi$ is uniform.
\qed

\section{The $\mmtm$: Stationary    Distributions}
\label{app::manhatdest}

Let's remind the key-property that allows to compute such stationary distributions.

\begin{obs} \label{obs::revers1}
 The $\mmtm$ is balanced, uniformly-selective and strongly-connected. So, from Theorems \ref{uniformity}
 and \ref{uniqueness}, the $\mmtm$ has a unique stationary distribution and it is the uniform one.
 Moreover, since the $\mmtm$ is  simple, the spatial and the destinational stationary distributions are given by
 Eq.s  \ref{pos} and \ref{eq::destsimple}, respectively.
 \end{obs}

\noindent Since the length of a path from a point $(i,j)$ to a point $(i',j')$ is $|i-i'|+|j-j'|$,
   with  some calculations,  we obtain
 
 
\begin{equation} 
|S(\tripset_\epsilon)|  \ = \ \frac{3}{(N^4-N^2)(4N-2)}\label{PROM}  
\end{equation}

\noindent Now let $\eta(u,v)$ be the number of paths starting from $u$ and visiting $v$. As for the
 Manhattan-MTM, we get

\begin{equation} \label{eq::ETA}
 \eta_\epsilon((i',j'),(i,j))=\left\{
\begin{array}{ll}
2N-i-j & \mbox{if $i'<i$ and $j'<j$}\\
i+j+2 & \mbox{if $i'>i$ and $j'>j$}\\
N+1-i+j & \mbox{if $i'<i$ and $j'>j$}\\
N+1+i-j & \mbox{if $i'>i$ and $j'<j$}\\
N^2-Nj & \mbox{if $i'=i$ and $j'<j$}\\
N^2-Ni & \mbox{if $i'<i$ and $j'=j$}\\
N+Nj & \mbox{if $i'=i$ and $j'>j$}\\
N+Ni & \mbox{if $i'>i$ and $j'=j$}\\
2N^2-2N+1 & \mbox{if $i'=i$ and $j'=j$}
\end{array}
\right.
\end{equation}

\noindent Thus
\begin{eqnarray} \label{eq::GAMMA} 
\Gamma_\epsilon (i,j)& =& \sum_{0\leq i'\leq N-1}\sum_{0\leq j'\leq N-1}\eta_\epsilon ((i',j'),(i,j))\nonumber\\
& =& (4N^2-6N+2)(i+j)-(4N-2)(i^2+j^2)+6N^2-8N+3
\end{eqnarray}

\noindent   From Eq.s \ref{pos}, \ref{eq::GAMMA}, \ref{PROM},   we get the spatial distribution
\begin{equation}\label{app:posm}
\spat_\epsilon (i,j)=\frac{3( (4N^2-6N+2)(i+j)-(4N-2)(i^2+j^2)+6N^2-8N+3)}{(N^4-N^2)(4N-2)}\cdot
\end{equation}

\noindent Our next   goal is to study the  Manhattan Random-Way Point  over grids of arbitrarily high
\emph{resolution}, i.e. for $\epsilon \rightarrow 0$. So, we will need to derive the \emph{probability-densitiy}
functions of the   stationary distribution. Let us compute the probability that an agent lies into a square of
center $(x,y)$ (where $x$ and $y$ are the Euclidean coordinates of a point in $V_{\epsilon}$) and side length $2
\delta$ w.r.t.  the   spatial distribution.

\[f_{\delta,\epsilon}(x,y)=\sum_{i\in\{\frac{N}{L}(x-\delta), \frac{N}{L}(x+\delta)\}}
\sum_{j\in\{\frac{N}{L}(y-\delta), \frac{N}{L}(y+\delta)\}}\pi_\epsilon(i,j)\] Then the     \emph{probability
density function} of the spatial distribution is given by
\begin{equation}\label{apx:rpd}
f(x,y)=\lim_{\delta\rightarrow 0}\lim_{\epsilon \rightarrow 0} \, \frac{1}{4\delta^2} \, f_{\delta,
\epsilon}(x,y) \, = \, \frac{3}{L^3}(x+y)-\frac{3}{L^4}({x}^2+{y}^2)
\end{equation}

\noindent 
   We now compute the  \emph{ agent-destination distribution}.   This    implies that  the
number of feasible paths visiting point $u$ and ending in point $v$ equals the number of feasible paths starting
from $u$ and visiting $v$, i.e.

\begin{equation} \label{gamma=eta}
\Gamma_{\epsilon,(i_0,j_0)}(i,j) =\eta_{\epsilon}((i,j),(i_0,j_0)) \end{equation}
 We now replace   Eq.s \ref{gamma=eta} and \ref{eq::GAMMA} into Eq. \ref{eq::destsimple} , and get the
 destination distribution

\begin{equation}\label{des:manh}
 \dest_{\epsilon,(i_0,j_0)}(i,j) = \frac{\eta_{\epsilon}((i,j),(i_0,j_0))}{\Gamma_\epsilon(i_0,j_0)}  \end{equation}

\noindent
 Let us now compute, the probability that an agent, visiting point $(x_0,y_0)$, has
  destination lying into the square of center $(x,y)$ and side
  length $2\delta$ (where $(x_0,y_0)$ and $(x,y)$  are the Euclidean
  coordinates of points in $V_\epsilon$). By definition
 of $\dest_{\epsilon,u}(v)$, it follows that

\[ f_{(x_0,y_0),\delta,\epsilon}(x,y)=\sum_{i\in\{\frac{N}{L}(x-\delta), \frac{N}{L}(x+\delta)\}}
\sum_{j\in\{\frac{N}{L}(y-\delta), \frac{N}{L}(y+\delta)\}}
\dest_{\epsilon,\left(\frac{N}{L}x_0,\frac{N}{L}y_0\right)}(i,j)   \] The probability density function of the
  destination distribution is
\[
f_{(x_0,y_0)}(x,y) = \lim_{\delta \rightarrow 0} \lim_{\epsilon \rightarrow 0} \frac{1}{4\delta^2}
f_{(x_0,y_0),\delta,\epsilon}(x,y)
\]
By combining  Eq.s \ref{des:manh}, \ref{eq::ETA},   \ref{eq::GAMMA},  and some calculations, we get

\begin{equation} \label{apx:eq::DENSITY}
 f_{(x_0,y_0)}(x,y) =\left\{
\begin{array}{ll}
  \frac {2L - x_0 - y_0}{4L(L(x_0+y_0) -(x_0^2 +y_0^2 ))} & \mbox{if $x<x_0$ and $y<y_0$}\\
\frac { x_0+y_0} {4L(L(x_0+y_0) -(x_0^2 +y_0^2 ))}     & \mbox{if $x>x_0$ and $y>y_0$}\\
     \frac { L-x_0 + y_0 }{4L(L(x_0+y_0) -(x_0^2 +y_0^2 ))} & \mbox{if $x<x_0$ and $y>y_0$}\\
  \frac { L + x_0 -y_0}{4L(L(x_0+y_0) -(x_0^2 +y_0^2 ))} & \mbox{if $x>x_0$ and $y<y_0$}\\
  + \infty& \mbox{if $x=x_0$ and $y<y_0$  (SOUTH Case)}\\
+ \infty  & \mbox{if $x<x_0$ and $y=y_0$  (WEST Case}\\
  + \infty & \mbox{if $x=x_0$ and $y>y_0$  (NORTH Case)}\\
 + \infty & \mbox{if $x>x_0$ and $y=y_0$  (EAST Case)}\\
 + \infty  & \mbox{if $x=x_0$ and $y=y_0$}
\end{array}
\right.
\end{equation}

\indent We now want to compute the stationary agent-destination distribution where the probability
 density function does not exist.
To this aim, for every \emph{segment} $\sc{s} \in \{$ SOUTH, WEST, NORTH,EAST $\}$,
 we evaluate the probability that an agent, visting point $(x_0,y_0)$, has destination
  lying into the subsegment of $\sc{s}$, having any given center and   length $2\delta$,
   \emph{conditioned} to the event that the destination belongs to  segment $\sc s$.
   From Eq. \ref{eq::destsimple}, we observe that, for any segment
$\sc s$,  it holds that

\[ \dest_u(v | v \in \sc s) = \frac{\Gamma_u(v)}{\sum_{w \in \sc s} \Gamma_u(w)} \]
We apply the above equation to each of the four cases and get, respectively,

\[
f^{\mbox{south}}_{(x_0,y_0),\delta, \epsilon} (y) =
 \frac{ \sum_{j \in \{(N/L)(y-\delta), (N/L)(y+\delta) \} }
 \eta_{\epsilon}( ((N/L)x_0,j), (  (N/L)x_0,(N/L)y_0))   }{ \sum_{j \in \{0, (N/L)y_0 \} }
 \eta_{\epsilon}( ((N/L)x_0,j), (  (N/L)x_0,(N/L)y_0))  }             \ \mbox{  for } y < y_0
\]

\[ f^{\mbox{west}}_{(x_0,y_0),\delta, \epsilon}(x) =
\frac {  \sum_{i \in \{(N/L)(x-\delta), (N/L)(x+\delta)\} }
 \eta_{\epsilon}( (i,(N/L)y_0), ((N/L)x_0,(N/L)y_0))   }{ \sum_{i \in \{0, (N/L)x_0 \} }
 \eta_{\epsilon}( (i,(N/L)y_0), ((N/L)x_0,(N/L)y_0))  }    \ \mbox{  for } x < x_0
\]

\[
f^{\mbox{north}}_{(x_0,y_0),\delta, \epsilon}(y) =
 \frac{ \sum_{j \in \{(N/L)(y-\delta), (N/L)(y+\delta)\} }
 \eta_{\epsilon}( ((N/L)x_0,j), (  (N/L)x_0,(N/L)y_0))   } { \sum_{j \in \{(N/L)y_0, N \} }
 \eta_{\epsilon}( ((N/L)x_0,j), (  (N/L)x_0,(N/L)y_0))  } \ \mbox{  for } y > y_0
\]

\[ f^{\mbox{east}}_{(x_0,y_0),\delta, \epsilon}(x) =
 \frac {  \sum_{i \in \{(N/L)(x-\delta), (N/L)(x+\delta)\} }
 \eta_{\epsilon}( (i,(N/L)y_0), ((N/L)x_0,(N/L)y_0))   }{ \sum_{i \in \{(N/L)x_0,N \} }
 \eta_{\epsilon}( (i,(N/L)y_0), ((N/L)x_0,(N/L)y_0))  }  \ \mbox{  for } x > x_0
\]

 \noindent
We consider the limits of the above four functions for $\epsilon \rightarrow 0$ and for $\delta \rightarrow 0$
and get the probability density functions of the destination distribution conditioned w.r.t. the four
segments.

\begin{equation} \label{eq::south}
f^{\mbox{south}}_{(x_0,y_0)}(y) =
  \lim_ {\delta\rightarrow 0} \lim_{\epsilon \rightarrow 0} \frac 1{2\delta}
   f^{\mbox{south}}_{(x_0,y_0),\delta, \epsilon}(y)
  = \frac{1}{y_0}
\end{equation}

\begin{equation} \label{eq::west}
f^{\mbox{west}}_{(x_0,y_0)}(x) =
  \lim_ {\delta\rightarrow 0} \lim_{\epsilon \rightarrow 0} \frac 1{2\delta} f^{\mbox{west}}_{(x_0,y_0),\delta,
   \epsilon}(x)
  = \frac{1}{x_0}
\end{equation}

\begin{equation} \label{eq::north}
f^{\mbox{north}}_{(x_0,y_0)}(y) =
  \lim_ {\delta\rightarrow 0} \lim_{\epsilon \rightarrow 0} \frac 1{2\delta} f^{\mbox{north}}_{(x_0,y_0),
  \delta, \epsilon}(y)
  = \frac{1}{ L-y_0}
\end{equation}

\begin{equation} \label{eq::east}
f^{\mbox{east}}_{(x_0,y_0)}(x) =
  \lim_ {\delta\rightarrow 0} \lim_{\epsilon \rightarrow 0} \frac 1{2\delta} f^{\mbox{east}}_{(x_0,y_0),\delta,
   \epsilon}(x)
  = \frac{1}{ L- x_0}
\end{equation}

\noindent From the above equations, it easy to see that the stationary destination distribution is uniform over
each of the four segments. We are now able to calculate the probability that an agent, visiting point $(x_0,y_0)$,
has destination in one of the four segments.

\[
\phi^{\mbox{south}}_{(x_0,y_0), \epsilon}  =
 \sum_{j \in \{ 0, (N/L)y_0 \} }
 \dest_{\epsilon,((N/L)x_0,(N/L)y_0)}((N/L)x_0,j)
\]

 \[ \phi^{\mbox{west}}_{(x_0,y_0), \epsilon}  =
  \sum_{i \in \{0, (N/L)x_0 \} }
 \dest_{\epsilon,((N/L)x_0,(N/L)y_0)}(i,(N/L)y_0)
\]

\[
\phi^{\mbox{north}}_{(x_0,y_0),  \epsilon}  =
  \sum_{j \in \{(N/L) y_0, N \} }
 \dest_{\epsilon,((N/L)x_0,(N/L)y_0)}((N/L)x_0,j)  \]

\[\phi^{\mbox{east}}_{(x_0,y_0),  \epsilon} =
  \sum_{i \in \{(N/L)x_0 ,  L \} }
 \dest_{\epsilon,((N/L)x_0,(N/L)y_0)}(i,(N/L)y_0)
\]

\noindent By taking the limits for $\epsilon \rightarrow 0$, we get, respectively,

\begin{equation}\label{eq::fisouth}
 \phi^{\mbox{south}}_{(x_0,y_0)} = \lim_{\epsilon \rightarrow 0}  \phi^{\mbox{south}}_{(x_0,y_0), \epsilon} =
\frac{y_0(L-y_0)}{4L(x_0+y_0) - 4(x_0^2 + y_0^2)}
 \end{equation}

\begin{equation}\label{eq::fiwest} \phi^{\mbox{west}}_{(x_0,y_0)} = \lim_{\epsilon \rightarrow 0}   \phi^{\mbox{west}}_{(x_0,y_0), \epsilon} =
\frac{x_0(L-x_0)}{4L(x_0+y_0) - 4(x_0^2 + y_0^2)}
 \end{equation}

\begin{equation}\label{eq::finorth}
\phi^{\mbox{north}}_{(x_0,y_0)} = \lim_{\epsilon \rightarrow 0}  \phi^{\mbox{north}}_{(x_0,y_0), \epsilon} =
\frac{y_0(L-y_0)}{4L(x_0+y_0) - 4(x_0^2 + y_0^2)}
 \end{equation}

\begin{equation}\label{eq::fieast}
\phi^{\mbox{east}}_{(x_0,y_0)} = \lim_{\epsilon \rightarrow 0}   \phi^{\mbox{east}}_{(x_0,y_0), \epsilon} =
\frac{x_0(L-x_0)}{4L(x_0+y_0) - 4(x_0^2 + y_0^2)}
 \end{equation}

\section{Modular MTM: Stationary Distributions} \label{apx::modular}

The next proposition provides the formulas for the stationary distributions of general 
modular MTM. Such formulas will be used to prove Proposition \ref{balanced_lego_formulas}.

\begin{prop}\label{lego_formulas}
Let $\mathfrak{R} = (\mathcal{B}, \mathcal{R})$ be a Route System such that the associated MTM 
$\mathcal{D}[\mathfrak{R}] = (\btripset{\mathfrak{R}}, \Psi[\mathfrak{R}])$ is strongly connected. Let $\sigma$ be
the stationary distribution of the Kernel of $\mathcal{D}[\mathfrak{R}]$. Then,
 the stationary spatial distribution $\mathfrak{s}$ of $\mathcal{D}[\mathfrak{R}]$ is, for every $u\in \mathcal{S}$,
{\small \[
\mathfrak{s}(u) =\; \frac{1}{\Lambda[\mathfrak{R}]} \sum_{B\in \mathcal{B}}
\frac{\#_{B, u}}{|B|} \sum_{w \in P(\btripset{\mathfrak{R}})} \frac{\sigma(w)}{\left|{\mathcal{R}}_{w}\right|} \sum_{R\in\mathcal{R}_w}\frac{\#_{R,B}}{|R|}
 \
\mbox{ with } 
\ 
\Lambda[\mathfrak{R}] \ =\; \sum_{B\in \mathcal{B}}
\frac{\#B}{|B|} \sum_{w \in P(\btripset{\mathfrak{R}})} \frac{\sigma(w)}{\left|\mathcal{R}_{w}\right|} 
\sum_{R\in\mathcal{R}_w}\frac{\#_{R,B}}{|R|}
\]
}The stationary destination distributions $\mathfrak{d}$ of $\mathcal{D}[\mathfrak{R}]$ are, for every $u, v\in \mathcal{S}$,
{\small \[
\mathfrak{d}_u(v) \;=\; \frac{1}{\mathfrak{s}(u)\Lambda[\mathfrak{R}]} \sum_{B\in \mathcal{B}}
\frac{\#_{B, u}}{|B|} \sum_{w \in P(\btripset{\mathfrak{R}})} \frac{\sigma(w)}{\left|{\mathcal{R}}_{w}\right|} \sum_{R\in\mathcal{R}_w\,\wedge R_{end}=v}\frac{\#_{R,B}}{|R|}
\]}
\end{prop}
 
\proof
Since $\mathcal{D}[\mathfrak{R}]$ is strongly connected, from Theorem~\ref{uniqueness}, $\mathcal{D}[\mathfrak{R}]$ has
e unique stationary distribution. Let $\pi$ be the stationary distribution of $\mathcal{D}[\mathfrak{R}]$ and let 
$\sigma$ be the (unique) stationary distribution of its Kernel. Firstly, we prove (i). From the definitions and 
Theorem~\ref{kernel_stationary}~(a), it holds that
\begin{eqnarray}
\mathfrak{s}(u)
& = & \nonumber
\sum_{T\in \btripset{\mathfrak{R}}_u} \#_{T, u} \pi(\langle T, 0\rangle)\\
& = & \nonumber
\sum_{T\in \btripset{\mathfrak{R}}_u} \#_{T, u} \frac{1}{\sum_{w\in P(\btripset{\mathfrak{R}})} \sigma(w)
\Lambda_{\Psi[\mathfrak{R}]}(w)}\sigma(T_{start}) \psi[\mathfrak{R}]_{T_{start}}(T)\\
& = & \label{lego_1}
 \frac{1}{\Lambda[\mathfrak{R}]} \sum_{T\in \btripset{\mathfrak{R}}_u} \#_{T, u} \sigma(T_{start}) \psi[\mathfrak{R}]_{T_{start}}(T)
\end{eqnarray}
To evaluate $\Lambda[\mathfrak{R}]$ we need the following claim
\begin{claim}\label{uno} $\sum_{T\in P} |T|=|\bpath{P}| \sum_{B\in P} \frac{\#B}{|B|}$.
\end{claim}
\proof
Let  $P=(B_1,B_2\ldots B_l)$, then \\
\begin{eqnarray}
\sum_{T\in P} |T| 
&=&\sum_{T\in B_1\cdot B_2\cdot \ldots B_l} |T| \nonumber \\
&=& \sum_{s_1\in B_1}\sum_{s_2\in B_2}\ldots \sum_{s_l\in B_l}(|s_1|+|s_2|+\ldots |s_l|)\nonumber \\
&=&\sum_{i=1}^l\sum_{s\in B_i}\frac{|s|\prod_{j=1}^l|B_j|}{|B_i|}\nonumber\\
&=&\sum_{i=1}^l\sum_{s\in B_i}\frac{|s|\cdot |\bpath{P}|}{|B_i|}\nonumber \\
&=&|\bpath{P}|\sum_{i=1}^l\frac{1}{|B_i|}\sum_{s\in B_i}|s|\nonumber\\
&=&|\bpath{P}|\sum_{i=1}^l\frac{1}{|B_i|}\#{B_i}\nonumber \\
&=&|\bpath{P}|\sum_{B\in P} \frac{\#B}{|B|} \nonumber
\end{eqnarray}
\qed

\begin{eqnarray}
\Lambda[\mathfrak{R}]
& = & \nonumber
\sum_{w\in P(\btripset{\mathfrak{R}})} \sigma(w) \Lambda_{\Psi[\mathfrak{R}]}(w)\\
& = & \nonumber
\sum_{w\in P(\btripset{\mathfrak{R}})} \sigma(w) \sum_{T\in \btripout{\mathfrak{R}}{w}} |T|\psi[\mathfrak{R}]_w(T)\\
& = & \nonumber
\sum_{w\in P(\btripset{\mathfrak{R}})} \sigma(w) \sum_{T\in \btripout{\mathfrak{R}}{w}} \frac{|T|}{\left|\mathcal{R}_w\right|}\sum_{R\in \mathcal{R}_w}\frac{1}{|R|}\sum_{P\in R}\frac{\#_{P,T}}{|\bpath{P}|}\\
& = & \nonumber
\sum_{w\in P(\btripset{\mathfrak{R}})} \frac{\sigma(w)}{\left|\mathcal{R}_w\right|} \sum_{R\in \mathcal{R}_w}\frac{1}{|R|}\sum_{P\in R}\frac{1}{|\bpath{P}|}
\sum_{T\in \btripout{\mathfrak{R}}{w}} |T|\cdot \#_{P,T}\\
& = & \nonumber
\sum_{w\in P(\btripset{\mathfrak{R}})} \frac{\sigma(w)}{\left|\mathcal{R}_w\right|} \sum_{R\in \mathcal{R}_w}\frac{1}{|R|}\sum_{P\in R}\frac{1}{|\bpath{P}|}
\sum_{T\in P} |T|\\
& = & \nonumber
\sum_{w\in P(\btripset{\mathfrak{R}})} \frac{\sigma(w)}{\left|\mathcal{R}_w\right|} \sum_{R\in \mathcal{R}_w}\frac{1}{|R|}\sum_{P\in R}\frac{|\bpath{P}|}{|\bpath{P}|}
\sum_{B\in P} \frac{\#B}{|B|}\quad \text{(by Claim \ref{uno}) }\\
& = & \nonumber
 \sum_{w\in P(\btripset{\mathfrak{R}})} \frac{\sigma(w)}{\left|\mathcal{R}_w\right|} \sum_{R\in \mathcal{R}_w}\frac{1}{|R|}\sum_{P\in R}
\sum_{B\in \mathcal{B}} \frac{\#_{P,B}\cdot \#B}{|B|}\\
& = & \nonumber
\sum_{B\in \mathcal{B}} \frac{ \#B}{|B|} \sum_{w\in P(\btripset{\mathfrak{R}})} \frac{\sigma(w)}{\left|\mathcal{R}_w\right|} \sum_{R\in \mathcal{R}_w}\frac{1}{|R|}\sum_{P\in R}\#_{P,B}
\\
& = & \nonumber
\sum_{B\in \mathcal{B}} \frac{ \#B}{|B|} \sum_{w\in P(\btripset{\mathfrak{R}})} \frac{\sigma(w)}{\left|\mathcal{R}_w\right|} \sum_{R\in \mathcal{R}_w}\frac{\#_{R,B}}{|R|}
\end{eqnarray}

Returning to Eq.~\ref{lego_1}, we firstly observe that the following claim holds 
\begin{claim}\label{due} $\sum_{T\in P} \#_{T,u}=|\bpath{P}| \sum_{B\in P} \frac{\#_{B,u}}{|B|}$.
\end{claim}
\proof Similar to the proof of Claim \ref{uno}. \qed

\noindent Thus  have that
\begin{eqnarray}
\mathfrak{s}(u)
& = & \nonumber
 \frac{1}{\Lambda[\mathfrak{R}]} \sum_{T\in \btripset{\mathfrak{R}}_u} \#_{T, u} \sigma(T_{start}) \psi[\mathfrak{R}]_{T_{start}}(T)\\
 & = & \nonumber
 \frac{1}{\Lambda[\mathfrak{R}]} \sum_{w \in P(\btripset{\mathfrak{R}})} 
  \sum_{T\in \btripout{\mathfrak{R}}{w}_u} \#_{T, u} \sigma(w) \psi[\mathfrak{R}]_{w}(T)\\
  & = & \nonumber
 \frac{1}{\Lambda[\mathfrak{R}]} \sum_{w \in P(\btripset{\mathfrak{R}})}  \sum_{T\in \btripout{\mathfrak{R}}{w}_u} \#_{T, u}\frac{\sigma(w)}{\left|\mathcal{R}_w\right|}
\sum_{R\in \mathcal{R}_w}\frac{1}{|R|}\sum_{P\in R}\frac{\#_{P,T}}{|\bpath{P}|}\\
   & = & \nonumber
 \frac{1}{\Lambda[\mathfrak{R}]} \sum_{w \in P(\btripset{\mathfrak{R}})}  \frac{\sigma(w)}{\left|\mathcal{R}_w\right|}
\sum_{R\in \mathcal{R}_w}\frac{1}{|R|}\sum_{P\in R}\frac{1}{|\bpath{P}|}\sum_{T\in \btripout{\mathfrak{R}}{w}_u} \#_{T, u}\cdot \#_{P,T}\\
    & = & \nonumber
 \frac{1}{\Lambda[\mathfrak{R}]} \sum_{w \in P(\btripset{\mathfrak{R}})}  \frac{\sigma(w)}{\left|\mathcal{R}_w\right|}
\sum_{R\in \mathcal{R}_w}\frac{1}{|R|}\sum_{P\in R}\frac{1}{|\bpath{P}|}\sum_{T\in P} \#_{T, u}\\
     & = & \nonumber
 \frac{1}{\Lambda[\mathfrak{R}]} \sum_{w \in P(\btripset{\mathfrak{R}})}  \frac{\sigma(w)}{\left|\mathcal{R}_w\right|}
\sum_{R\in \mathcal{R}_w}\frac{1}{|R|}\sum_{P\in R}\frac{|\bpath{P}|}{|\bpath{P}|}\sum_{B\in P} \frac{\#_{B,u}}{|B|}\quad \text{(by Claim \ref{due}) }\\
& = & \nonumber
 \frac{1}{\Lambda[\mathfrak{R}]} \sum_{w \in P(\btripset{\mathfrak{R}})}  \frac{\sigma(w)}{\left|\mathcal{R}_w\right|}
\sum_{R\in \mathcal{R}_w}\frac{1}{|R|}\sum_{B\in \mathcal{B}}\frac{\#_{R,B}\cdot \#_{B,u}}{|B|}\\
 & = & \nonumber
\frac{1}{\Lambda[\mathfrak{R}]} \sum_{B\in \mathcal{B}}
\frac{\#_{B, u}}{|B|} \sum_{w \in P(\btripset{\mathfrak{R}})} \frac{\sigma(w)}{\left|{\mathcal{R}}_{w}\right|} \sum_{R\in\mathcal{R}_w}\frac{\#_{R,B}}{|R|}
\end{eqnarray}
This proves formula (i).

Now consider  formula (ii) for the stationary  destination distributions. From the definitions,
Theorem~\ref{kernel_stationary}~(a), and very similar calculations as those used for the stationary spatial 
distribution, we have  
\begin{eqnarray}
\mathfrak{d}_u(v) 
& = & \nonumber
\frac{1}{\mathfrak{s}(u)} \sum_{T\in \btripset{\mathfrak{R}}_u \wedge T_{end} = v} \#_{T, u}\cdot \pi(\langle T, 0\rangle)\\
& = & \nonumber
\frac{1}{\mathfrak{s}(u)\Lambda[\mathfrak{R}]}\sum_{T\in \btripset{\mathfrak{R}}_u \wedge T_{end} = v} \#_{T, u} 
\sigma(T_{start}) \psi[\mathfrak{R}]_{T_{start}}(T)\\
& = & \nonumber
\frac{1}{\mathfrak{s}(u)\Lambda[\mathfrak{R}]} \sum_{w \in P(\btripset{\mathfrak{R}})} \sum_{T\in \btripout{\mathfrak{R}}{w}_u 
\wedge T_{end} = v} \#_{T, u} \sigma(w) \psi[\mathfrak{R}]_{w}(T)\\
& = & \nonumber
\frac{1}{\mathfrak{s}(u)\Lambda[\mathfrak{R}]} \sum_{w \in P(\btripset{\mathfrak{R}})} \sum_{T\in \btripout{\mathfrak{R}}{w}_u 
\wedge T_{end} = v} \#_{T, u}  \frac{\sigma(w)}{\left|\mathcal{R}_w\right|}\sum_{R\in \mathcal{R}_w}\frac{1}{|R|}\sum_{P\in R}\frac{\#_{P,T}}{|\bpath{P}|}\\
& = & \nonumber
\frac{1}{\mathfrak{s}(u)\Lambda[\mathfrak{R}]} \sum_{w \in P(\btripset{\mathfrak{R}})}  \frac{\sigma(w)}{\left|\mathcal{R}_w\right|}\sum_{R\in \mathcal{R}_w}\frac{1}{|R|}\sum_{P\in R}\frac{1}{|\bpath{P}|}\sum_{T\in \btripout{\mathfrak{R}}{w}_u 
\wedge T_{end} = v} \#_{T, u} \cdot \#_{P,T}\\
& = & \label{lego_x}
\frac{1}{\mathfrak{s}(u)\Lambda[\mathfrak{R}]} \sum_{w \in P(\btripset{\mathfrak{R}})}  \frac{\sigma(w)}{\left|\mathcal{R}_w\right|}\sum_{R\in \mathcal{R}_w}\frac{1}{|R|}\sum_{P\in R}\frac{1}{|\bpath{P}|}\sum_{T\in P \wedge T_{end} = v} \#_{T, u} 
\end{eqnarray}
Notice that if $R\in \mathcal{R}_w$ is such that $R_{end}\neq v$ then 
$$\sum_{P\in R}\frac{1}{|\bpath{P}|}\sum_{T\in P \wedge T_{end} = v} \#_{T, u} =0$$
Thus, returning to Eq.~\ref{lego_x}, we have

\begin{eqnarray}
\mathfrak{d}_u(v) 
& = & \nonumber
\frac{1}{\mathfrak{s}(u)\Lambda[\mathfrak{R}]} \sum_{w \in P(\btripset{\mathfrak{R}})}  \frac{\sigma(w)}{\left|\mathcal{R}_w\right|}\sum_{R\in \mathcal{R}_w}\frac{1}{|R|}\sum_{P\in R}\frac{1}{|\bpath{P}|}\sum_{T\in P \wedge T_{end} = v} \#_{T, u} \\
& = & \nonumber
\frac{1}{\mathfrak{s}(u)\Lambda[\mathfrak{R}]} \sum_{w \in P(\btripset{\mathfrak{R}})}  \frac{\sigma(w)}{\left|\mathcal{R}_w\right|}\sum_{R\in \mathcal{R}_w\wedge R_{end} = v}\frac{1}{|R|}\sum_{P\in R}\frac{1}{|\bpath{P}|}\sum_{T\in P } \#_{T, u} \\
& = & \nonumber
\frac{1}{\mathfrak{s}(u)\Lambda[\mathfrak{R}]} \sum_{w \in P(\btripset{\mathfrak{R}})}  \frac{\sigma(w)}{\left|\mathcal{R}_w\right|}\sum_{R\in \mathcal{R}_w\wedge R_{end} = v}\frac{1}{|R|}\sum_{P\in R}\frac{|\bpath{P}|}{|\bpath{P}|}\sum_{B\in P } \frac{\#_{B, u}}{|B|} \quad \text{(by Claim \ref{due}) }\\
& = & \nonumber
 \frac{1}{\mathfrak{s}(u)\Lambda[\mathfrak{R}]} \sum_{w \in P(\btripset{\mathfrak{R}})}  \frac{\sigma(w)}{\left|\mathcal{R}_w\right|}
\sum_{R\in \mathcal{R}_w\wedge R_{end} = v}\frac{1}{|R|}\sum_{B\in \mathcal{B}}\frac{\#_{R,B}\cdot \#_{B,u}}{|B|}\\
 & = & \nonumber
\frac{1}{\mathfrak{s}(u)\Lambda[\mathfrak{R}]} \sum_{B\in \mathcal{B}}
\frac{\#_{B, u}}{|B|} \sum_{w \in P(\btripset{\mathfrak{R}})} \frac{\sigma(w)}{\left|{\mathcal{R}}_{w}\right|} \sum_{R\in\mathcal{R}_w\wedge R_{end} = v}\frac{\#_{R,B}}{|R|}
\end{eqnarray}
\qed

\subsection{Proof of Proposition \ref{balanced_lego_formulas}}


We need the following preliminary lemmas.

\begin{lemma}\label{balance_lemma}
Let $\mathcal{D} = (\tripset, \Psi)$ be a MTM such that   a function $f : P(\tripset)\rightarrow \mathbb{R}$
exists  satisfying
the following properties:
\begin{enumerate}[(i)]
\item
\[
\sum_{u\in P(\tripset)} f(u) > 0 \qquad\text{and}\qquad \forall u\in P(\tripset)\quad f(u) \geqslant 0
\]
\item
\[
\forall u\in P(\tripset)\quad \sum_{T\in \tripin{u}} f(T_{start}) \psi_{T_{start}}(T) \;=\; f(u)
\]
\end{enumerate}
Then, the following map
\[
\forall u\in P(\tripset)\quad \sigma(u) \;=\; \frac{f(u)}{\sum_{v\in P(\tripset)} f(v)}
\]
is a stationary distribution of the Kernel of $\mathcal{D}$.
\end{lemma}
\proof
From Property (a), $\sigma$ is a probability distribution over $P(\tripset)$. It holds that, for every $u\in P(\tripset)$,
\begin{eqnarray*}
\sum_{v\in P(\tripset)} \sigma(v)\Prob{v\rightarrow u}
& = &
\sum_{v\in P(\tripset)} \sigma(v) \sum_{T\in \trips{v}{u}} \psi_v(T)\\
& = &
\sum_{v\in P(\tripset)} \frac{f(v)}{\sum_{w\in P(\tripset)} f(w)} \sum_{T\in \trips{v}{u}} \psi_v(T)\\
& = &
\frac{1}{\sum_{w\in P(\tripset)} f(w)} \sum_{v\in P(\tripset)}\sum_{T\in \trips{v}{u}} f(v)\psi_v(T)\\
& = &
\frac{1}{\sum_{w\in P(\tripset)} f(w)}  \sum_{T\in \tripin{u}} f(T_{start})\psi_{T_{start}}(T)\\
& = &
\frac{1}{\sum_{w\in P(\tripset)} f(w)} f(u) \quad \text{(by Property (b))}\\
& = &
\sigma(u)
\end{eqnarray*}
\qed

\begin{lemma}\label{balanced_bs_lemma}
Let $\mathfrak{R} = (\mathcal{B}, \mathcal{R})$ be a balanced Route System. Then, the map $\sigma$ so defined
\[
\forall u\in P(\btripset{\mathfrak{R}})\quad \sigma(u) \;=\; \frac{\left|\mathcal{R}_{u}\right|}{|\mathcal{R}|} 
\]
is a stationary distribution of the Kernel of the associated MTM
$\mathcal{D}[\mathfrak{R}]$.
\end{lemma}
\proof
Consider the function $f : P(\btripset{\mathfrak{R}}) \rightarrow \mathbb{R}$ such that $f(u) = \left|\mathcal{R}_{u}\right|$,
for every $u\in P(\btripset{\mathfrak{R}})$. Clearly, $f$ satisfies Property~(a) of Lemma~\ref{balance_lemma}.
To verify that it also satisfies Property~(b) w.r.t. the MTM $\mathcal{D}[\mathfrak{R}]$, consider the following
\begin{eqnarray*}
 \sum_{T\in \btripin{\mathfrak{R}}{u}} f(T_{start}) \psi[\mathfrak{R}]_{T_{start}}(T)
 & = &
 \sum_{T\in \btripin{\mathfrak{R}}{u}}  \frac{f(T_{start})}{\left|\mathcal{R}_{T_{start}}\right|}\sum_{R\in \mathcal{R}_{T_{start}}}\frac{1}{|R|}\sum_{P\in R}\frac{\#_{P,T}}{|\bpath{P}|}\\
 & = &
 \sum_{v\in P(\btripset{\mathfrak{R}}) } \sum_{T\in \btripset{\mathfrak{R}}_{(v,u)} }  \sum_{R\in \mathcal{R}_v}\frac{1}{|R|}\sum_{P\in R}\frac{\#_{P,T}}{|\bpath{P}|}\\ 
 & = &
 \sum_{v\in P(\btripset{\mathfrak{R}}) }  \sum_{R\in \mathcal{R}_v}\frac{1}{|R|}\sum_{P\in R}\frac{1}{|\bpath{P}|}\sum_{T\in \btripset{\mathfrak{R}}_{(v,u)} }\#_{P,T} \\ 
 & = &
 \sum_{v\in P(\btripset{\mathfrak{R}}) }  \sum_{R\in \mathcal{R}_{v,u}}\frac{1}{|R|}\sum_{P\in R}\frac{1}{|\bpath{P}|}\sum_{T\in \btripset{\mathfrak{R}}_{(v,u)} }\#_{P,T} \\ 
& = & 
\sum_{v\in P(\btripset{\mathfrak{R}}) }  \sum_{R\in \mathcal{R}_{v,u}}\frac{1}{|R|}\sum_{P\in R}\frac{1}{|\bpath{P}|}|\bpath{P}| \\ 
& = & 
\sum_{v\in P(\btripset{\mathfrak{R}}) }  \left| \{ R\in \mathcal{R}_{v} | \mathcal{R}_{end}=u \}\right| \\ 
& = & 
\left|\{R\in \mathcal{R}|R_{end}=u\}\right| \\ 
& = & 
 \left| \mathcal{R}_u\right|  \quad \text{(since $\mathfrak{R}$ is balanced)}\\
 & = &
 f(u)
\end{eqnarray*}
Hence, we can apply Lemma~\ref{balance_lemma} and so we obtain that the map  $\sigma$, defined as
\[
\sigma(u) \;=\; \frac{f(u)}{\sum_{w\in P(\tripset)} f(w)} \;=\; \frac{\left|\mathcal{R}_{u}\right|}{|\mathcal{R}|} 
\]
is a stationary distribution of the Kernel of the MTM $\mathcal{D}[\mathfrak{R}]$.
\qed

\smallskip
\noindent
\textbf{Proof of Proposition \ref{balanced_lego_formulas}.}
 Since $\mathfrak{R}$ is balanced and $\mathcal{D}[\mathfrak{R}]$ is strongly connected, from Lemma~\ref{balanced_bs_lemma}, 
the unique stationary distribution of the Kernel of $\mathcal{D}[\mathfrak{R}]$ is the following
\[
\forall u\in P(\btripset{\mathfrak{R}})\quad \sigma(u) \;=\; \frac{\left|\mathcal{R}_{u}\right|}{|\mathcal{R}|} 
\]
We firstly consider the spatial stationary distribution. From Proposition~\ref{lego_formulas}~(i) and the above expression
for $\sigma$ we obtain
\begin{eqnarray}
\mathfrak{s}(u) 
& = & \nonumber
\frac{1}{\Lambda[\mathfrak{R}]} \sum_{B\in \mathcal{B}}
\frac{\#_{B, u}}{|B|} \sum_{w \in P(\btripset{\mathfrak{R}})} \frac{\sigma(w)}{\left|{\mathcal{R}}_{w}\right|} \sum_{R\in\mathcal{R}_w}\frac{\#_{R,B}}{|R|}\\
& = & \nonumber
\frac{1}{\Lambda[\mathfrak{R}]} \sum_{B\in \mathcal{B}}
\frac{\#_{B, u}}{|B|} \sum_{w \in P(\btripset{\mathfrak{R}})} \frac{1}{\left|{\mathcal{R}}\right|} \sum_{R\in\mathcal{R}_w}\frac{\#_{R,B}}{|R|}\\
& = & \label{blego_1}
\frac{1}{\Lambda[\mathfrak{R}]}  \frac{1}{\left|{\mathcal{R}}\right|} \sum_{B\in \mathcal{B}}
\frac{\#_{B, u}}{|B|} \sum_{R\in\mathcal{R}}\frac{\#_{R,B}}{|R|}
\end{eqnarray}
Moreover, it holds that
\begin{eqnarray}
\Lambda[\mathfrak{R}] 
& = & \nonumber
\sum_{B\in \mathcal{B}}
\frac{\#B}{|B|} \sum_{w \in P(\btripset{\mathfrak{R}})} \frac{\sigma(w)}{\left|\mathcal{R}_{w}\right|} 
\sum_{R\in\mathcal{R}_w}\frac{\#_{R,B}}{|R|}\\
& = & \nonumber
\sum_{B\in \mathcal{B}}
\frac{\#B}{|B|} \sum_{w \in P(\btripset{\mathfrak{R}})} \frac{1}{\left|\mathcal{R}\right|} 
\sum_{R\in\mathcal{R}_w}\frac{\#_{R,B}}{|R|}\\
& = & \nonumber
\frac{1}{\left|\mathcal{R}\right|} \sum_{B\in \mathcal{B}}
\frac{\#B}{|B|} 
\sum_{R\in\mathcal{R}}\frac{\#_{R,B}}{|R|}\\
& = &\label{blego_2}
\frac{1}{\left|\mathcal{R}\right|}  \Lambda_{\mathrm{b}}[\mathfrak{R}]
\end{eqnarray}
By combining Equalities~\ref{blego_1} and \ref{blego_2}, we obtain
\[
\mathfrak{s}(u) \;=\; \frac{1}{\Lambda_{\mathrm{b}}[\mathfrak{R}]} \sum_{B\in \mathcal{B}}
\frac{\#_{B, u}}{|B|}\sum_{R\in\mathcal{R}}\frac{\#_{R,B}}{|R|}
\]

The proof of formula (ii) is straightforward and it is very similar to the one of formula (i).
\qed

\section{The DownTown Formulas} \label{apx:boundlecounting}
 In this section, we give the spatial distribution formulas for some further kinds of cell of the DownTown Model.
 
 \noindent
  {\bf - Start-bundles.}   We consider a  starting-bundle $\bstop{S}{++}$ into a horizontal block (i.e. even $i$ and odd $j$).
  Observe that the number of bundle-paths containing this bundle equlas the number of parking cells located into blocks 
  whose horizontal 
  coordinate is not smaller than $j$  (but the cells into block $(i,j)$). Then, 
  
  \begin{equation} \label{eq:++}
   \sbstop{S}{++}  \ = \ m ((n+1)(n-j+1) - 2)          \end{equation}
  
  \noindent
  Similarly, for a  starting-bundle $\bstop{S}{++}$ into a horizontal block,  we get 
  
   \begin{equation} \label{eq:--}
    \sbstop{S}{--}  \ = \ m ((n+1)(j+1) - 2)          \end{equation}
   
   \noindent
   As for $\sbstop{S}{+-}$, it is equal to number of parking cells located into blocks 
  whose horizontal coordinate is less than $j$. So, we get
   \begin{equation} \label{eq:+-}
    \sbstop{S}{+-}  \ = \  m ((n+1) j  - 1)       
       \end{equation}

\noindent
Similarly, for a  starting-bundle $\bstop{S}{-+}$ into a horizontal block,  we get 
   
  \begin{equation}  \label{eq:-+}
   \sbstop{S}{-+}  \ = \  m ((n+1) (n-j) - 1)   
  \end{equation}

\noindent
As for starting-bundles   into  vertical blocks (i.e. odd $i$ and even  $j$), we can get the correct formulas
for $\sbstop{S}{**}$ by replacing index $j$ with $i$.

\noindent
{\bf - End-bundles.}   We consider an  end-bundle $\bstop{E}{++}$ into a horizontal block (i.e. even $i$ and odd $j$).
  Observe that the number of bundle-paths containing this bundle equals the number of parking cells located into blocks 
  whose horizontal 
  coordinate is   smaller than $j$ plus  the number of  positive parking stripes    located into blocks whose horizontal coordinate is
  $j$ (but  the positive cells into block $(i,j)$).
Then, 
  
  \begin{equation} \label{eq:e++}
   \sbstop{E}{++}  \ = \ m ((n+1)j +  n/2 -1 )          \end{equation}

\noindent
  Similarly, for a  starting-bundle $\bstop{E}{--}$ into a horizontal block   we get 
  
   \begin{equation}  \label{eq:e--}
    \sbstop{S}{--}  \ = \ m ((n+1)(n-j) +  n/2 -1 )        \end{equation}

 \noindent
    Similarly, for the  other two cases   we get
   \begin{equation} \label{eq:e+-}
    \sbstop{E}{+-}  \ = \  m ((n+1) (n-j) + n/2 - 1)       
       \end{equation}

 \begin{equation} \label{eq:e-+}
    \sbstop{E}{-+}  \ = \  m ((n+1) j  + n/2 - 1)       
       \end{equation}

\noindent
As for ending-bundles   into  vertical blocks (i.e. odd $i$ and even  $j$), we can get the correct formulas
for $\sbstop{S}{**}$ by replacing index $j$ with $i$.

\noindent
{\bf -  Transit-bundles.}  By applying counting arguments, we get the correct formulas for transit-bundles.
Define   function 

\begin{equation*} 
 \eta(i,j) = \left \{  \begin{array}{ll}
  m^2(n-j+1) (nj -(j+1))  +  m^2(n-j-1) (nj +2j + n -i ) &     \mbox{$i$ even,   $j$  odd  with }  \\ 
                                                     &  i   \notin \{0, n \}  \\
    m^2(n-j+1) (nj -\frac 32 (j+1))  +  m^2(n-j-1) (nj +2j + n  ) &     \mbox{$i=0$  and    $j$  odd }  \\ 
  m^2(n-j+1) (nj -\frac {n+1}2 (j+1))  +  m^2(n-j-1) (nj +2j   ) &     \mbox{$i=n$  and    $j$  odd }  \\                                                       
0   &     \mbox{ otherwise }  \end{array}
\right.
\end{equation*}

\noindent
Then, we get 
\begin{equation}  \label{bt+}
 \sbtrans{+}(i,j) = \left \{  \begin{array}{ll}
  \eta(i,j)  &     \mbox{$i$ even,   $j$  odd    }  \\ 
           \eta(j,i)  &    \mbox{$i$ odd,   $j$  even    } \\ 
   0   &     \mbox{ otherwise }  \end{array}
\right.
\end{equation}
 
 By exploiting symmetric properties, from the above formula  we   get 

\begin{eqnarray}
\sbtrans{-}(i,j) &  =  & \sbtrans{+}(n-j,i)  \label{bt-}  
\end{eqnarray}

\noindent
{\bf - Cross-bundles.} 
By applying counting arguments, we get the correct formulas for cross-bundles.

\begin{equation} \label{crh++}
 \sbcross{H}{++}(i,j) = \left \{  \begin{array}{ll}
  m^2(n^2-n + (2n-1)j - (n-1)i - 2ij ) &     \mbox{$i$ and $j$  even  with }  \\ 
                                                     &  i \neq n  \mbox{ and }  j \neq 0  \\
0  &     \mbox{ otherwise }  \end{array}
\right.
\end{equation}

By exploiting symmetric properties, from the above formula  we   get 

\begin{eqnarray}
\sbcross{V}{+-}(i,j) &  =  & \sbcross{H}{++}(n-j,i)  \label{crv+-}  \\
\sbcross{H}{--}(i,j)  & = & \sbcross{H}{++}(n-i,n-j)  \label{crh--}  \\
\sbcross{V}{-+}(i,j) & =  &\sbcross{H}{++}(j,n-i)  \label{crv-+}    
\end{eqnarray}

\noindent
It   holds that 

\begin{equation} \label{crH+}
 \sbcross{H}{+}(i,j) = \left \{  \begin{array}{ll}
  m^2 (n-j) (2nj -i + j)  &     \mbox{ $i$ and $j$  even  with }  \\ 
                                                     &  i,j \notin \{0, n \}    \\
  m^2 (n-j) (2nj  + j/2)  &     \mbox{ $i=0$ and $j$  even  with }  \\ 
                                                     &  j \notin \{0, n \}    \\
  m^2 (n-j) (\frac 32 j(n+1) -n)     &     \mbox{ $i=n$ and $j$  even  with }  \\ 
                                                     &  j \notin \{0, n \} \\
  0    &     \mbox{ otherwise }  \end{array}
\right.
\end{equation}

\noindent
By exploiting symmetric properties, from the above formula  we   get 

\begin{eqnarray}
\sbcross{V}{+}(i,j) &  =  & \sbcross{H}{+}(n-j,i)  \label{crv+}  \\
\sbcross{H}{-}(i,j)  & = & \sbcross{H}{+}(n-i,n-j)  \label{crh-}  \\
\sbcross{V}{-}(i,j) & =  &\sbcross{H}{+}(j,n-i)  \label{crv-}    
\end{eqnarray}

\noindent
It   holds that 

\begin{equation} \label{crH+-}
 \sbcross{H}{+-}(i,j) = \left \{  \begin{array}{ll}
  m^2 (2ij + i + j)  &     \mbox{ $i$ and $j$  even  with }  \\ 
                                                     &  j \neq 0 \mbox{ and }  i \notin \{0, n \} \\
  m^2 ( nj + n + j )  &     \mbox{ $i=n$ and $j$  even  with }  \\ 
                                                     &  j \neq  0     \\
  0    &     \mbox{ otherwise }  \end{array}
\right.
\end{equation}

\noindent
By exploiting symmetric properties, from the above formula  we   get 

\begin{eqnarray}
\sbcross{V}{++}(i,j) &  =  & \sbcross{H}{+-}(n-j,i)  \label{crv++}  \\
\sbcross{H}{-+}(i,j)  & = & \sbcross{H}{+-}(n-i,n-j)  \label{crh-+}  \\
\sbcross{V}{--}(i,j) & =  &\sbcross{H}{+-}(j,n-i)  \label{crv--}    
\end{eqnarray}

\medskip
\noindent
\textbf{Computing the Downt-Town Spatial Distribution: Relevant locations.}
 We now  use  the above values  to compute
the stationary spatial distribution given by Prop. \ref{balanced_lego_formulas}. 
We   give  such formulas for some kind of cells  (the others can be obtained by the same counting arguments).
Let $\Lambda = \Lambda_{\mathrm{b}}[\mathfrak{R^D}] / m^2$  be the normalization constant
 with $\Lambda_{\mathrm{b}}[\mathfrak{R^D}]$ defined in Prop \ref{balanced_lego_formulas}.  

 \noindent
\textbf{-  [Transit Blocks].}
Let $u$ be a   transit  cell of index  $k$ (with any  $k \in \{1, \ldots , m\}$) in the \emph{positive transit stripe} of the horizontal block  $(i,j)$  (i.e. $i$ even and $j$ odd).   
  (with $i \notin \{0,n\}$).
Then 
\begin{eqnarray} 
\mathfrak{s}(u) & = & \frac{\slk(k)}{\Lambda_{\mathrm{b}}[\mathfrak{R^D}]} \left(   \sbtrans{+}(i,j)   +\sum_{h=1}^{k}\sigma\mbox{-}B_{S,h}^{++}(i,j) +\sbstop{S}{+-}  +  \sum_{h=k}^{m}\sigma\mbox{-}B_{E,h}^{++}(i,j)  +\sbstop{E}{+-} \right)\nonumber 
\end{eqnarray}
Thus, since $\sigma\mbox{-}B_{S,h}^{*,*}(i,j)$ and $\sigma\mbox{-}B_{E,h}^{++}(i,j)$ do not depend on $h$, we have
\begin{eqnarray} 
\mathfrak{s}(u) & = & \frac{\slk(k)}{\Lambda_{\mathrm{b}}[\mathfrak{R^D}]} \left(   \sbtrans{+}(i,j)   +k\cdot\sbstop{S}{++}  +\sbstop{S}{+-}  + (m-k+1)\sbstop{E}{++}   +\sbstop{E}{+-} \right)\nonumber\\
& = & \frac{\slk(k)}{\Lambda_{\mathrm{b}}[\mathfrak{R^D}]} m( m(n-j+1)(nj-(j+1)+m(n-j-1)(nj+2j+n-i)+k((n+1)(n-j+1)-2)+\nonumber\\
&& (n+1)j-1+(m-k+1)((n+1)j+\frac{n}{2}-1)+(n+1)(n-j)+\frac{n}{2}-1)\nonumber\\
&=&\frac{\slk(k)}{\Lambda}\left(a(i,j)+\frac{k}{m}b(j)+\frac{1}{m}c(j)\right)
\end{eqnarray} 
where
\begin{eqnarray}
a(i,j) &=& (n-j)(2nj+j+n-i-1)+(n-2)j-\frac{n}{2}+i-2\nonumber\\
b(j)& = &n(n+1)+\frac{n}{2}-2(n+1)j\nonumber\\
c(j)& =& (n+1)(n+j)+n-3\nonumber
\end{eqnarray}
 \noindent

\textbf{-  [Parking Cells].}
Let $u$ be a  parking cell of index $k$  (with any  $k \in \{1, \ldots , m\}$) in the \emph{positive parking stripe} of the horizontal block  $(i,j)$  (i.e. $i$ even and $j$ odd).
Then 
\begin{eqnarray}
 \mathfrak{s}(u) & = & \frac{\wait}{\Lambda_{\mathrm{b}}[\mathfrak{R^D}]} \left(   \sbstop{S}{++}   +\sbstop{S}{+-}    +  \sbstop{E}{++}  +\sbstop{E}{+-}  \right)\nonumber \\
&=& \frac{\wait}{\Lambda_{\mathrm{b}}[\mathfrak{R^D}]} m(  (n+1)(n-j+1) - 2 + (n+1) j  - 1 +  (n+1)j +  n/2 -1  + \nonumber\\
&& +(n+1) (n-j) + n/2 - 1)\nonumber\\
&=&\frac{\wait}{\Lambda}\frac{2n^2+4n-4}{m}
\end{eqnarray}
\noindent
\textbf{-  [Cross-Ways].}
Let $u$ be a  crossing cell  $(+,-)$   in  the crossing block  $(i,j)$  (i.e. $i,j$ even ) with $i,j\not\in\{0,n\}$
Then 
\
\begin{eqnarray}
 \mathfrak{s}(u) & = & \frac{\crc}{\Lambda_{\mathrm{b}}[\mathfrak{R^D}]} \left(   \sbcross{H}{+}(i,j)  +\sbcross{H}{+-}(i,j)  + \sbcross{V}{-}(i,j)  +\sbcross{V}{-+}(i,j) \right)\nonumber \\
&=& \frac{\crc}{\Lambda_{\mathrm{b}}[\mathfrak{R^D}]} m^2(  (n-j)(2nj-i+j) +2ij+i+j+ 
i(2n(n-i)-j+n-i)+\nonumber\\
& & +n^2-n+(2n-2j-1)(n-i)-(n-1)j)\nonumber\\
&=&\frac{\crc}{\Lambda}\left(n^2(2j+2i+3)-2n(j^2+i^2+j+i+1) -(i-j)^2+2(ij+j+i)\right)
\end{eqnarray}

\end{small}
\end{document}

\end{thebibliography}


\begin{thebibliography}{10}


 
  \bibitem{KPL08} 
  J. Koberstein,  H. Peters, and N.  Luttenberger.
 \newblock  Graph-based mobility model for urban areas fueled with real world datasets.
\newblock  In Proc. of the 1st \emph{SIMUTOOLS '08},  1--8, 2008.
  


\bibitem{AF99}
D.~Aldous and J.~Fill.
\newblock {\em Reversible Markov Chains and Random Walks on Graphs}.
\newblock http://stat-www.berkeley.edu/users/aldous/RWG/book.html, 2002.





\bibitem{CBD02}
T.~Camp, J.~Boleng, and V.~Davies.
\newblock A survey of mobility models for ad hoc network research.
\newblock {\em Wireless Communication and Mobile Computing}, 2(5), 483--502,
  2002.

\bibitem{BB87} F. Baccelli and P. Br\'emaud. \newblock
\emph{Palm Probabilities and Stationary Queues}. \newblock Springer-Verlag, 1987.


\bibitem{BRS03} C. Bettstetter, G. Resta, and P. Santi.
\newblock The Node Distribution of the Random Waypoint Mobility Model for Wireless Ad Hoc Networks.
\newblock  \emph{IEEE Transactions on
Mobile Computing}, 2, 257--269, 2003.



\bibitem{CNB04}
T.~Camp, W.~Navidi, and N.~Bauer.
\newblock Improving the accuracy of random waypoint simulations through
  steady-state initialization.
\newblock In Proc. {\em  of 15th Int. Conf. on Modelling and Simulation},
  319--326, 2004.

 
\bibitem{CMPS09}
A.~Clementi, A.~Monti, F.~Pasquale, and R.~Silvestri.
\newblock Information spreading in stationary markovian evolving graphs.
\newblock In Proc. {\em  of the 23rd IEEE IPDPS},  1--12, 2009.

\bibitem{CPS09}
A.~Clementi,   F.~Pasquale, and R.~Silvestri.
\newblock MANETS: High mobility can make up for low transmission power.
\newblock  In Proc. of  \emph{ICALP'09}, LNCS 5556, 387--398, 2009.

\bibitem{CMS10b}
A.~Clementi, A.~Monti,   and R.~Silvestri.
\newblock Flooding over Manhattan.
\newblock Submitted paper, 2010.

\bibitem{CDMRV09}
P.~Crescenzi, M.~Di Ianni, A.~Marino, G.~Rossi , and P.~Vocca. \newblock Spatial Node Distribution of Manhattan
Path Based Random Waypoint Mobility Models with Applications.
\newblock
In Proc. of \emph{SIROCCO'09}, LNCS,  2009.





\bibitem{DPSW08}
J.~Diaz, X.~Perez, M.J. Serna, and N.C. Wormald.
\newblock Walkers on the cycle and the grid.
\newblock {\em SIAM J. Discrete Math.}, 22(2), 747--775, 2008.

\bibitem{DMP08}
J.~Diaz, D.~Mitsche, and X.~Perez-Gimenez.
\newblock On the connectivity of dynamic random geometric graphs.
\newblock In Proc. {\em  of the 19th   ACM-SIAM 
   SODA'08},  601--610, 2008.

 
\bibitem{GS97}
\newblock C. Grinstead and J. L. Snell
\newblock {\em Introduction to Probability}.
\newblock American Mathematical Society, 1997.
\newblock Online version in \verb"http://www.dartmouth.edu/~chance/teaching_aids/books_articles/" \verb"probability_book/book.html"

 
  
\bibitem{L06}
J-Y. Le Boudec and M. Vojnovic.
\newblock The Random Trip Model: Stability, Stationary Regime, and Perfect Simulation.
\newblock \emph{IEEE/ACM Transaction on Networking}, 16(6), 1153--1166,  2006.

\bibitem{L07}
J-Y. Le Boudec. \newblock Understanding the simulation of mobility models with Palm calculus.
\newblock
\emph{Performance Evaluation}, 64, 126--147, 2007.

\bibitem{LV05}
J-Y. Le Boudec and M.~Vojnovic.
\newblock Perfect simulation and the stationarity of a class of mobility
  models.
\newblock In Proc.  {\em  of the 24th IEEE INFOCOM},    2743--2754, 2005.




\bibitem{W_BF}
\newblock  \verb"http://en.wikipedia.org/wiki/Brouwer_fixed_point_theorem"

\end{thebibliography}
\end{document}

1.)  Representing  the \emph{exact} geometric agent position is not so relevant since it can be  approximated by 
      a good cell partition of the geometric space and, then, recording the \emph{cell}\footnote{\emph{Cells} are   the basic elements of the   support
      space: they are also called \emph{points} or \emph{positions}.} the  agent lies on, at every  time. \\
       2.) Time is a real continuous variable; however, storing the agent position at every real value $t$ is again not necessary:
            we can observe and record  agent positions  only at discrete time steps, forgetting what happens in between. \\
           The required approximation   can  be  tuned by fixing the resolution of the cell partition and the
            value of the time-unit.